\documentclass[a4paper,11pt]{article}
\pdfoutput=1
\usepackage{graphicx}
\usepackage{jcappub}
\usepackage[T1]{fontenc}
\usepackage{lipsum}
\usepackage{natbib}
\usepackage{amsbsy}
\usepackage{soul}
\usepackage{symate}
\usepackage{journals}
\usepackage{units}
\usepackage{multirow}
\usepackage{color}
\usepackage{comment}
\usepackage{booktabs}
\usepackage{subcaption}
\usepackage{xspace}
\usepackage{placeins}
\usepackage{rotating,tabularx}
\usepackage{array,booktabs}
\usepackage[font=small,labelfont=bf]{caption}
\usepackage{pifont}
\graphicspath{{./Pictures/}}
\newcommand{\xmark}{\ding{55}}
\newcommand{\be}{\begin{equation}}
\newcommand{\ee}{\end{equation}}
\newcommand{\bea}{\begin{eqnarray}}
\newcommand{\eea}{\end{eqnarray}}
\newcommand{\nn}{\nonumber}
\def\camb{{\sc{CAMB}}\xspace}
\def\halofit{{\sc{Halofit}}\xspace}
\def\hmcode{{\sc{HMCODE}}\xspace}
\def\cosmicemu{{\sc{CosmicEMU}}\xspace}
\def\NGhalofit{{\sc{NGenHalofit}}\xspace}

\def\etab{\eta_{\mathrm b}}
\def\logMc{\log M_{\mathrm c}}
\def\kMpc{\, h \, {\rm Mpc}^{-1}}
\newcommand{\eq}[1]{eq.~(\ref{#1})}

\title{\boldmath On the degeneracy between baryon feedback and massive neutrinos as probed by matter clustering and weak lensing}

\author[a,b]{Gabriele Parimbelli}
\author[a,b,c,d]{, Matteo Viel}
\author[c,b]{, Emiliano Sefusatti}

\affiliation[a]{SISSA - International School for Advanced Studies, Via Bonomea 265, 34136 Trieste, Italy}
\affiliation[b]{INFN - National Institute for Nuclear Physics, Via Valerio 2, 34127 Trieste, Italy}
\affiliation[c]{INAF - Osservatorio Astronomico di Trieste, Via Tiepolo 11, 34131 Trieste, Italy}
\affiliation[d]{IFPU - Institute for Fundamental Physics of the Universe, Via Beirut 2, 34151 Trieste, Italy}

\emailAdd{gparimbe@sissa.it}
\emailAdd{viel@sissa.it}
\emailAdd{emiliano.sefusatti@inaf.it}

\abstract{ Massive neutrinos, due to their free streaming, produce a suppression in the matter power spectrum at
  intermediate and small scales which could be probed by galaxy clustering
  and/or weak lensing observables. This effect happens at scales that
  are also influenced by baryon feedback, i.e. galactic winds or
  Active Galactic Nuclei (AGN) feedback, which in realistic
  hydrodynamic simulations has also been shown to produce a
  suppression of power. Leaving aside, for the moment, the complex issue of galaxy bias, we focus here on matter clustering and
  tomographic weak lensing, we investigate the possible degeneracy between
  baryon feedback and neutrinos showing that it is not likely to degrade significantly the measurement
  of neutrino mass in future surveys. To do so, we generate mock data
  sets and fit them using the Markov Chain Monte Carlo (MCMC)
  technique and explore degeneracies between feedback parameters and
  neutrino mass.  We model baryon feedback through fitting functions,
  while massive neutrinos are accounted for, also in the non-linear
  regime, using \textsc{Halofit} calibrated against accurate N-body
  neutrino simulations.  In the error budget, we include the
  uncertainty in the modelling of non-linearities. For both matter clustering and weak lensing, we always recover the
  input neutrino mass within $\sim0.25\sigma$ confidence level.
  Finally, we also take into account the intrinsic alignment effect in
  the weak lensing mock data.  Even in this case, we are able to
  recover the right parameters: in particular, we find a significant  degeneracy pattern between $M_\nu$ and the intrinsic alignment parameter
  $A_\mathrm{IA}$.}

\begin{document}
\maketitle
\flushbottom


\section{Introduction}
\label{sec:intro}

In the upcoming years galaxy surveys like
\textit{Euclid}\footnote{\url{https://www.euclid-ec.org/}}, the Large
Synoptic Survey Telescope
(\textsc{LSST})\footnote{\url{https://www.lsst.org/}}, the Dark Energy
Spectroscopic Instrument
(\textsc{DESI})\footnote{\url{http://desi.lbl.gov/}} and the Square
Kilometer Array
(\textsc{SKA})\footnote{\url{https://www.skatelescope.org/}} will
become operative.  Indeed, some of these ambitious projects are
already happening, see for example
\textsc{DES}\footnote{\url{https://www.darkenergysurvey.org/}}
\cite{DES_1yr+18, DES_lensing+17}.  Thanks to these probes we will be
able to study the evolution of the Universe through cosmic ages, using
as observables galaxy clustering (baryon acoustic
oscillations, BAO, and redshift-space distortions, RSD) and weak lensing, which will
be measured with unprecedented accuracy.  Such improvements will allow
to constrain better the cosmological parameters and assess possible
deviations from the standard flat $\Lambda$-Cold Dark Matter
($\Lambda$CDM) paradigm.  In particular, these new experiments will
almost certainly be able to measure for the first time the total
neutrino mass, $M_\nu$, which is known to suppress the growth of
structures at small scales.
A lower bound of $M_\nu = 0.056$ eV is obtained by particle physics experiments from neutrino oscillations (see e.g. \cite{LP_massive_nu+06}); on the other hand, cosmology so far has been able only to place either upper limits \cite{Giusarma+16} or marginal preference \cite{Beutler+14, Battye+14, DiValentino+17} for a non-zero total neutrino mass.
To date, the most stringent constraint comes from combining Planck \cite{Planck+15} with BOSS Lyman-$\alpha$ forest data, providing $M_\nu < 0.12$ eV at 95\% confidence level \cite{PD-lya-nu+15}.

To achieve the goal of measuring cosmological parameters with percent accuracy, a huge effort must be carried out in the theoretical modelling of the observables and of the systematic errors that may affect our predictions for RSD and galaxy bias in clustering observations and shape-noise in shear surveys.
One of the most important aspects when dealing with massive neutrinos is the fact that their effect on the matter power spectrum may mimic what goes under the name of \textit{baryon feedback}.
With baryon feedback we mean here the set of astrophysical processes able to modify the matter distribution on the relatively large/intermediate scales comparable to the halo sizes.
Such processes include Active Galactic Nuclei (AGN) feedback, galactic winds and hot X-ray emitting gas physics.
Since the observational constraints are still poor, their description usually relies on hydrodynamical N-body simulations \cite{OWLs+10, Van_Daalen-feedback+11, Semboloni+11, Mummery-fb_vs_nu+17}.
It is also true that baryon effects are underestimated in simulations: while observations have shown that even massive halos are missing significant amounts of gas \cite{Sun+09, Lovisari+15, Eckert+16}, simulations tend to overpredict the baryon fraction in clusters. BAHAMAS and FABLE simulations make exception, as they are calibrated on the observed baryon fractions \cite{BAHAMAS+17,FABLE+18}.
What is found is that, despite these mechanisms are different in nature, their net effect on the matter power spectrum is similar: we expect a damping starting from $k\sim 0.5\kMpc$ with a maximum suppression of 10-25 percent around $k \sim 2 \ \kMpc$ \cite{Schneider+15}.
In the past few years, several analytical approaches have been proposed to model such suppression: some authors propose fitting functions whose parameters depend on the feedback model implemented in the simulation \cite{HD_feedback+15, Chisari-feedback+18}, others treat feedback as the consequence of a modification of the Navarro-Frenk-White (NFW) density profile in dark matter halos \cite{Mead-b+15, Schneider+15}.

Previous and current works have always treated baryon feedback as a nuisance to the cosmological signal \cite{DES_1yr+18, Koh-KiDS+17},
without quantifying a possible bias on the estimate of $M_\nu$.
In this paper we show that the degeneracy between baryon feedback and massive neutrinos is not significant and the two effects can indeed be disentangled.
One of the main reason why this can be done is that, besides involving different ranges of scales, the redshift evolution of the two phenomena are rather different from each other: thus, we expect that  performing a tomographic analysis will be the best way to achieve our goal.

Our work is based on the formalism developed in Ref.~\cite{Audren+12}, where a Markov Chain Monte Carlo (MCMC) method to forecast errors on cosmological parameters for a \textit{Euclid}-like survey combined with Planck is used.
A more recent paper \cite{Sprenger+18} has carried out the same analysis including also 21cm galaxy clustering and intensity mapping as will be probed by SKA.
MCMC has been shown to return more realistic results than a simple Fisher matrix analysis, which has well-known limits when exploring the cosmological parameters {\em full} likelihood function.

In this work we consider as our ``observables'' the 3-D, {\em matter} power spectrum and the cosmic shear power spectrum. Clearly the former is not a real observable and, in fact, in this paper we limit ourselves to a study of the degeneracy between neutrino mass and baryon feedback {\em at the level of the matter distribution}, leaving any complication due to galaxy bias (and redshift-space distortions) for a future work.
Furthermore, for clustering we apply the so-called \textit{cold dark matter prescription} \cite{Ichiki+11,nuLCDM1+13,nuLCDM2+13,nuLCDM3+13}: in massive neutrino cosmologies, the relevant field for the description of galaxy clustering is the cold dark matter plus baryon (cdm+b) one rather than the total matter, since we can recover the expected constant linear bias at large scales only w.r.t. the former. In our set-up, therefore, the observable for clustering is the cdm+b power spectrum.

Weak lensing is one of the most promising tools for cosmology for the next decade: despite resulting from an integration along the line-of-sight, it probes the deep non-linear regime, as non-linearities arise already at relatively small multipoles ($\ell\sim 100$) \cite{Takada_WL+04}.
The latest results from the KiDS survey \cite{Koh-KiDS+17} and DES \cite{DES_lensing+17} {show some tensions between each other.
In addition, KiDS data seem to underpredict the overall amplitude of primordial fluctuations with respect to Planck \cite{Lensing_is_low+16, Problems_KiDS+17}.
Ref. \cite{McCarthy+17} has demonstrated that baryon feedback alone is not enough to reconcile the tension in the $\Omega_\mathrm{m}-\sigma_8$ plane and a non-minimal neutrino mass  can resolve it.
Therefore a detailed study of the possible degeneracies between the two effects becomes of primary importance.

We incorporate non-linearities using the latest version of the \textsc{Halofit} subroutine implemented in \textsc{Camb} \cite{Mead-b+15, Mead-nu+16}\footnote{\url{https://camb.info/}}.
These fitting formulae can reproduce the cold matter power spectrum with an accuracy better than $\sim5\%$ even in the deeply non-linear regime ($k\lesssim 10 \kMpc$).
With this approach we essentially assume that any prediction adopted in the analysis of future data set is the direct results of investigations based on N-body simulations and we therefore include as theoretical uncertainties those affecting numerical methods. 
How this and other sources of error are computed can be found explicitly in Section \ref{sec:method}.

This paper is organized as follows: in Section \ref{sec:theory} we briefly summarise the theory behind the observables we employ and introduce the equations that we use; in Section \ref{sec:method} we present the details of our analysis; in Section \ref{sec:results} we show our results; finally, in Section \ref{sec:discussion} we draw our conclusions.

Throughout this work we assume a flat \LCDM model with one single massive neutrino species with parameters $\Omega_\mathrm b = 0.0486$, $\Omega_\mathrm m = \Omega_\mathrm{cdm}+\Omega_\nu + \Omega_\mathrm b= 0.3089$, $h = 0.6774$, $A_\mathrm s = 2.14\times 10^{-9}$, $n_\mathrm s = 0.9667$, i.e. the best-fit values found by the Planck Collaboration (2015) \cite{Planck+15}.

\section{Theory}
\label{sec:theory}

In this Section we briefly summarise the theoretical framework and the definitions of the cdm+b and the shear power spectra.
Then we describe the baryon feedback model of Ref. \cite{Schneider+15} that we will use throughout this work.


\subsection{Observables: cdm+b power spectrum and cosmic shear}
\label{sec:observables}

The large-scale structure of the Universe is the result of the growth
of small density perturbations that evolved through cosmic ages.  In
massive neutrino cosmologies, at the redshifts relevant for
large-scale structure observations, we can identify two contributions
to the total matter density $\rho_\textrm m$ given by the cold matter
(including baryons), $\rho_\textrm c$, and neutrinos,
$\rho_\nu$. Total matter perturbations can then be written as 
\be
\delta_\textrm m=(1-f_\nu)\,\delta_\textrm c+f_\nu\,\delta_\nu\,, 
\ee
where $f_\nu\equiv\Omega_\nu/\Omega_m$ is the fraction of the neutrino
contribution to the total matter density.  From now on, the subscript
`c' will label the cdm+b fluid, while `m' will denote total matter
quantities. Numerical investigations \cite{nuLCDM1+13, nuLCDM2+13,
  nuLCDM3+13, DEMNUni+16, Paco-Neutrinos+17, Ruggeri+18}, along with
earlier theoretical descriptions of the spherical collapse in the
massive neutrino scenario \cite{Ichiki+11}, have shown that halo
formation is driven by the cold matter component alone. This
assumption allows to recover universality for the halo mass function
and halo bias, otherwise apparently lost in a description based on
total matter perturbations. The quantity of interest for halo (and
galaxy) clustering is therefore the cold matter power spectrum,
$P_\textrm{cc}(k)$. In this work we will focus on this quantity
leaving, for the moment, aside the actual observable, i.e.~the galaxy
power spectrum. The aim of this work, in fact, is not to present the
most realistic errors on the total neutrino mass achievable by future
surveys but rather to quantitatively address the putative degeneracies
between the suppressions of power induced by feedback effects and
neutrinos \textit{at the level of matter perturbations alone} .

Our second quantity of interest is the shear power spectrum from gravitational lensing \cite{Bartelmann_WL+01, Hoek_WL+08}. We will work in the weak regime, where distortions of the shapes of galaxies are much smaller that their intrinsic ellipticity. Intervening mass distorts the image of the source galaxy in both size and ellipticity, but in this regime the power spectra of the two components (convergence and shear) are statistically equivalent. The lensing effect depends on the gravitational potential along the line-of-sight, directly related, via Poisson equation, to the {\em total} matter perturbations. Our predictions for this observable will then be derived in terms of the total matter power spectrum $P_{\rm mm}(k)$.

Dividing source galaxies into $N$ redshift bins, i.e.~performing a tomographic analysis, allows to improve the constraints on cosmological parameters \cite{Takada_WL+04}, as this will result in $N(N+1)/2$ nearly independent observables. We assume the flat-sky Limber's approximation, which are valid for small angles or, equivalently, for high $\ell$ values ($\ell \gtrsim 10-20$) \cite{Kilbinger-full-sky+17}. The cosmic shear power spectrum then reads
\be
C^{(ij)}(\ell) = \int_0^\infty \de z \ \frac{c}{H(z)} \ \frac{W^{(i)}(z) \ W^{(j)}(z)}{\chi^2(z)} \ P_\mathrm{mm}\left(k=\frac{\ell}{\chi(z)}, z \right)\,,
\label{eq:shear}
\ee
where $P_\mathrm{mm}(k,z)$ is the non-linear total matter power spectrum and $\chi(z)$ is the comoving distance to redshift $z$.
$W^{(i)}$ is the window function describing the (normalised) distribution of sources $n_\mathrm s(z)$ in the redshift bin $[z_i, z_{i+1}]$, given by
\be
W^{(i)}(z) = \frac{3}{2} \ \Omega_\mathrm{m}  \left(\frac{H_0}{c}\right)^2 (1+z) \ \chi(z) \int_{\min{(z, z_i)}}^{z_{i+1}}  \de x \ n_\mathrm s(x) \ \frac{\chi(x)-\chi(z)}{\chi(x)}
\ee
and is therefore a measurement of the lensing efficiency in the $i$-th tomographic bin.


\subsection{Massive neutrinos}
\label{sec:feedback}

Massive neutrinos affect the cold and total matter power spectra (and in turn the cosmic shear one) both in the linear and non-linear regimes. The large thermal velocities that neutrinos possess at their decoupling prevent them from clustering in regions smaller than the free-streaming scale (e.g. \cite{LP_massive_nu+06}):
\begin{equation}
k_{\textrm{fs}} = 0.82 \ \frac{E(z)}{(1+z)^2} \ \frac{M_\nu}{1 \unit{\ eV}}  \kMpc,
\end{equation}
where $M_\nu$ is the sum of the neutrino masses and $E(z)= H(z)/H_0$. This results in a suppression of the linear cdm+b power spectrum $P_\mathrm{cc}(k)$ and total matter power spectrum $P_\mathrm{mm}(k)$ that can be quantified, in the small-scale limit, as \cite{LP_massive_nu+06, DEMNUni+16}
\be
\frac{\Delta P_\mathrm{cc}^\mathrm L(k)}{P_\mathrm{cc}^\mathrm L(k)} \approx -6 \ f_\nu,\qquad\frac{\Delta P_\mathrm{mm}^\mathrm L(k)}{P_\mathrm{mm}^\mathrm L(k)} \approx -8 \ f_\nu\,,
\ee
as long as $f_\nu  \lesssim 0.07$ \cite{LP_massive_nu+06}.  Clearly we expect a larger suppression in the total matter power spectrum since this is given by the combination 
\be
P_{\rm mm}(k)=(1-f_\nu)^2\,P_{\rm cc}(k)+2\,f_\nu\,(1-f_\nu)\,P_{\rm c\nu}(k)+f_\nu^2\,P_{\rm \nu\nu}(k)\,,
\ee
where the cross cold matter-neutrinos power spectrum $P_{\rm c\nu}(k)$ and neutrinos power spectrum  $P_{\rm \nu\nu}(k)$  rapidly decay for $k>k_{\rm fs}$.

At lower redshift, as the temperature drops, neutrinos become non-relativistic and eventually fall into dark matter potential wells. This ``neutrino drag'' relieves the small-scale suppression \cite{Bird-HALOFIT+12}, so that plotting the ratio between the power spectra in a massive and massless neutrino cosmology returns the well-known spoon-shape curve around $k \sim 1 \kMpc$ when compared to a massless neutrino model with the same amplitude in the large-scale perturbations  (see the continuous lines in the top left panel of Figure \ref{fig:Pk_nu_feed}). Quantitatively, in the non-linear regime, the suppression becomes of order $\Delta P/P \approx -10 \ f_\nu$ with a stronger scale-dependence \cite{brandbyge08, vielhaehneltspringel2010}.

Because of the integration of \eq{eq:shear}, the suppression of the matter power spectrum translates into a suppression in the shear power spectrum that affects almost all multipoles, with a milder dependence on scale.

\subsection{Baryon feedback}

In addition to massive neutrinos, baryonic feedback processes, comprising violent events such as supernova explosions and the accretion onto the central black hole in AGNs, are also  responsible for a small-scale drop in power. Theoretical predictions for such effects are both poorly constrained by observations and inevitably affected by large systematic uncertainties, due to the difficulty in properly capture baryonic physics in numerical simulations.
In general, baryon feedback is expected to damp the matter power spectrum by up to 25\% at scales of $k\sim 2 \kMpc$ \cite{Schneider+15}, but the uncertainty caused by different AGN feedback models could reach 50\% for scales $k\lesssim 1 \kMpc$ \cite{HD_feedback+15}.

Nevertheless, in the last few years several analytical descriptions, relying on fits to numerical simulations, have been proposed.
In this paper we will use the \textit{baryon correction model} (BCM) by Ref. \cite{Schneider+15}.
As opposed to other similar proposals (see for instance \cite{HD_feedback+15, Chisari-feedback+18}) this model has the advantage of employing parameters with a well-established physical meaning.
The BCM assumes that X-ray emitting gas, AGN activity and more in general the complex intracluster physics smoothly modify the NFW profile of the dark matter halo.
This modification reflects on the matter power spectrum in a way that can be explained by a fitting function with only three free parameters.
The BCM is obtained from a set of hydrodynamical simulations that incorporate AGN feedback but not other mechanisms such as galactic winds, which could produce different scale and redshift dependencies for the suppression.
Although the suppression of power is model-dependent (i.e. depends on different sets of parameters), the shape of the predicted damping, as obtained from simulations, is similar for most of the feedback mechanisms. 
This is the main motivation to relax the priors on the BCM feedback parameters even outside their physical range in order to be conservative and probe a wider range of feedback-induced suppressions.

The BCM fitting function describes the ratio between the total matter power spectrum accounting for baryon feedback to the power spectrum of the dark-matter-only (dmo) scenario and reads
\begin{equation}
F_\mathrm{bf}(k, z|M_c, \eta_\mathrm b, z_\mathrm c) \equiv \frac{P_{\mathrm{feed}}(k)}{P_{\mathrm{dmo}}(k)} = \left\{\frac{B(z)}{1+(k/k_g)^3} + \left[1-B(z)\right]\right\} \ S(k),
\label{eq:feedback_function}
\end{equation}
where
\begin{equation}
B(z) = \frac{0.105 \log{\left(\frac{M_c}{\unit{M_\odot}/h}\right)} - 1.27}{1+(z/z_\mathrm c)^{2.5}} \,,
\label{eq:feedback_function_1}
\end{equation}
for  $M_c \geq 10^{12} \unit{\ M_\odot}/h$ and zero otherwise,
\begin{equation}
k_g(z) = 0.7 \ [1-B(z)]^4 \ \eta_\mathrm b^{-1.6} \kMpc\,,
\label{eq:feedback_function_2}
\end{equation}
while the term outside the bracket is the stellar component of the central galaxy,
\begin{equation}
S(k) = 1+\left(\frac{k}{55 \kMpc}\right)^2\,.
\label{eq:feedback_function_3}
\end{equation}

The expressions above depend on three parameters: $M_c$, $\eta_\mathrm b$ and $z_\mathrm c$. The critical mass $M_c$ is related to the bound gas fraction in a halo. Hydrodynamical simulations show that part of this gas is ejected and such ejection is stronger in low mass halos. So we expect low mass halos to be almost completely stripped from their gas.  In this picture, $M_c$ represents the typical halo mass scale below which most of the gas is ejected.
This parameter sets the prominence of the suppression: the higher $M_c$, the smaller $P_{\mathrm{feed}}(k)$ will be.
The parameter $\eta_\mathrm b$ controls the maximum scale (minimum $k$) at which the suppression becomes relevant.
Such parameter is related to the ejected gas fraction: it may be viewed as the ratio between the thermal velocity of the gas in the intracluster medium and the halo escape velocity. As such, the higher $\eta_\mathrm b$, the more the suppression occurs at larger scales.
Finally, the last parameter $z_\mathrm c$ accounts for the time dependence of the suppression, which is growing with decreasing redshift as the signal is dominated by larger and larger halos.

Ref. \cite{Schneider+15} tested the BCM on the hydrodynamical simulations by Ref. \cite{Jing+05} which include radiative cooling and star formation but no AGN feedback, obtaining a best fit of $M_c \sim 2 \times 10^{12} \ M_\odot/h$ and $\eta_\mathrm b \sim 1.0$.
Such low value for $M_c$ means that only very low mass halos are completely emptied of their gas, in agreement with the lack of AGN feedback set in the simulations.
BCM was also applied to the OWLs simulations \cite{Van_Daalen-feedback+11} obtaining values of $M_c \sim 5 \times 10^{14} \ M_\odot/h$ and $\eta_\mathrm b \sim 0.4$, indicating a high AGN activity. The systematic error affecting \eq{eq:feedback_function} is of order $2-3$\% at all scales up to $k\lesssim 10 \kMpc$.
This uncertainty will be included in the error on the \textsc{Halofit} formulae that we will introduce in the next Section.

Figures \ref{fig:Pk_nu_feed} and \ref{fig:Cl_nu_feed} show separately the effects of massive neutrinos and baryon feedback, respectively on the matter power spectrum and the shear power spectrum.
The top left panel of Figure \ref{fig:Pk_nu_feed} shows how increasing the neutrino mass damps more and more the matter power spectrum (both at linear and non-linear level).
A key point is that the scale at which the suppression occurs is almost constant, being it only weakly-dependent on $M_\nu$, and it is much larger than the scales involved by baryon feedback (see the other three panels).

Figure \ref{fig:Cl_nu_feed} shows the same effect but on cosmic shear. 
One can see that feedback only affects high multipoles ($\ell\gtrsim 80$), while massive neutrinos damp the shear spectrum even at low multipoles.
For neutrino masses greater than 0.3 eV, the suppression is so high that, in order for baryon feedback to mimic it, all halos with mass smaller than $\sim 10^{14} \ M_\odot/h$ should expel their gas, implying an extraordinarily strong AGN activity.

\begin{figure}[!t]
	 \makebox[1\textwidth][c]
		{
			\includegraphics[width=1.05\textwidth]{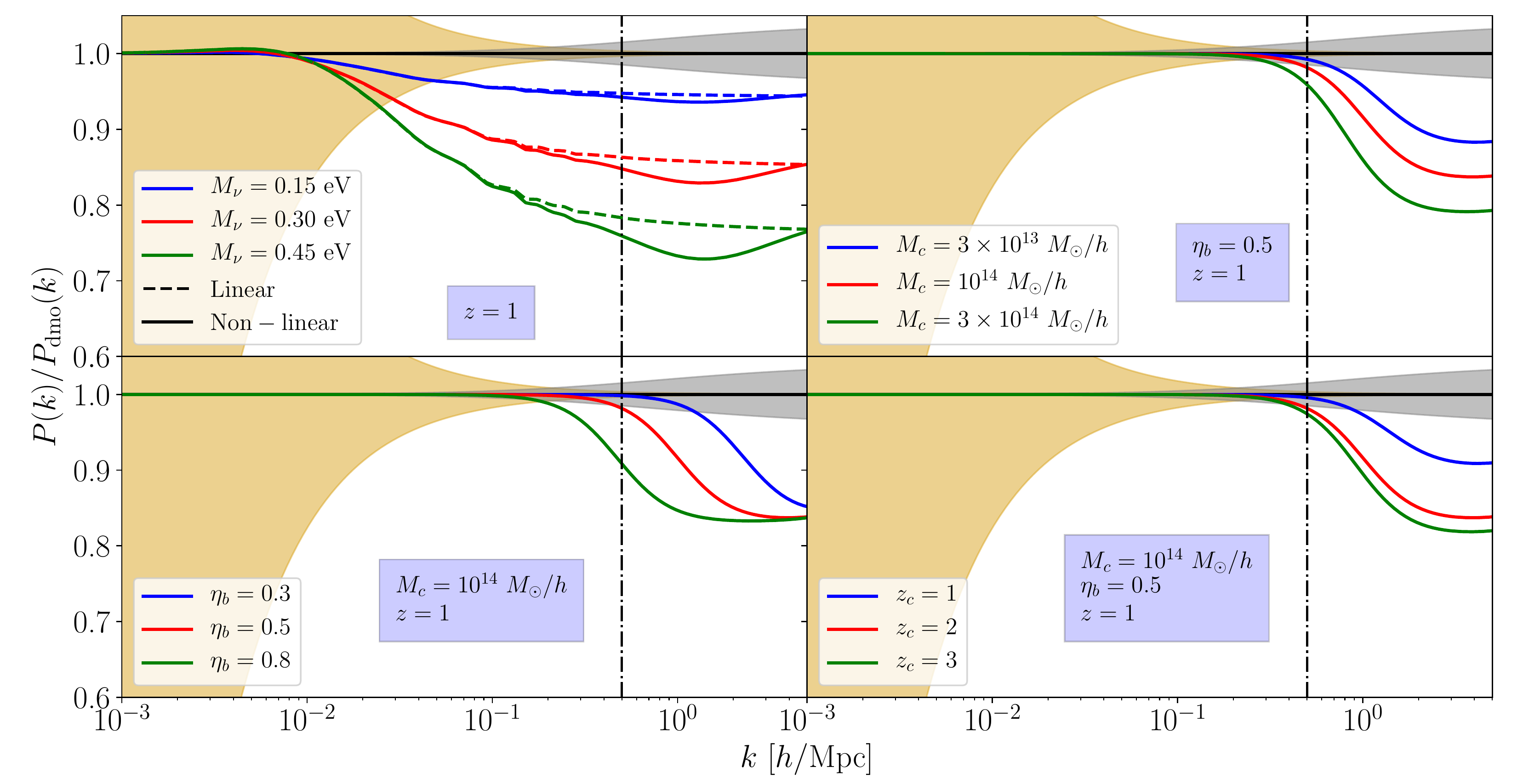}
   		  		 } 
\caption{The four panels show the effect of the neutrino mass and of the three different feedback parameters of the BCM on the matter power spectrum at $z=1$ (approximately the median redshift of future surveys). All the ratios are taken with respect to a cdm+b power spectrum model (linear w.r.t. the linear, non-linear w.r.t. the non-linear) in a cosmology with minimum-allowed neutrino mass $M_\nu = 0.056$ eV and no baryon feedback. In the top left panel the neutrino mass is varied while the ratio in both linear and non-linear regime are shown. The top right panel shows the effect of increasing $\log M_c$, in the bottom left panel we change the parameter $\eta_\mathrm b$, while in the bottom right we display how the redshift parameter affects the feedback fitting function. The gold shaded areas represent cosmic variance for a survey like in Ref. \cite{Audren+12} in a redshift bin of $\Delta z = 0.1$ centered at $\overline z=1$. The grey shaded areas represent the theoretical uncertainty on the matter power spectrum due to the \textsc{Halofit} fitting formulae, \eq{eq:halofit_error}. In all the panels a vertical line at $k = 0.5 \kMpc$ is drawn, to mark the maximum $k$ at which our analysis is extended.}
 \label{fig:Pk_nu_feed}
\end{figure}

\begin{figure}[!t]
	 \makebox[1\textwidth][c]
		{
			\includegraphics[width=1.05\textwidth]{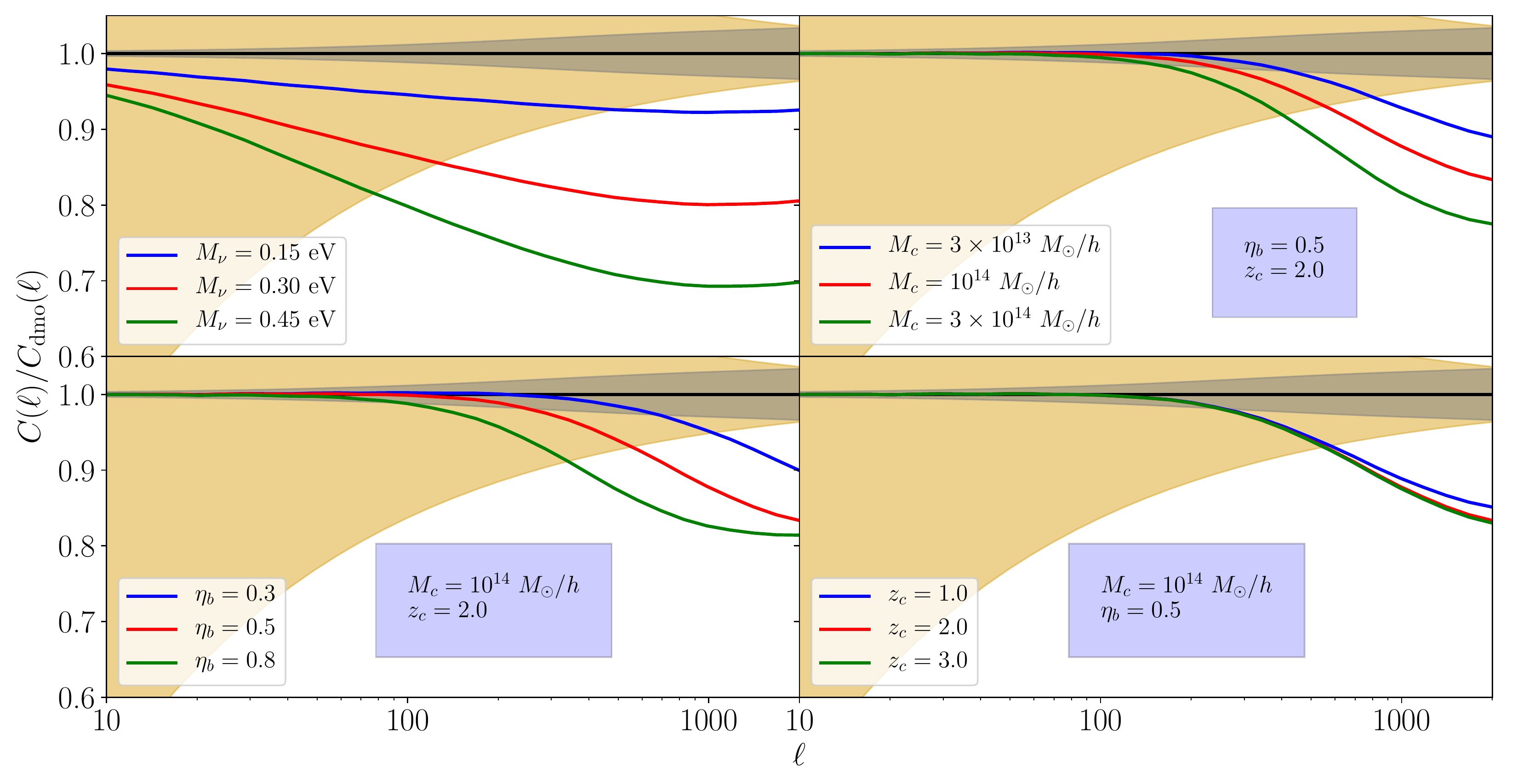}
   		  		 } 
\caption{This picture is the same of Figure \ref{fig:Pk_nu_feed} but here the cosmic shear power spectrum is shown, with a source distribution like in Ref. \cite{Audren+12}, a sky coverage of $f_\textrm{sky} = 0.375$ and with all galaxies in a single bin, i.e. no tomography has been performed.}
 \label{fig:Cl_nu_feed}
\end{figure}

\section{Method}
\label{sec:method}

The goal of this paper is to investigate possible degeneracies between the effects of massive neutrinos and baryon feedback on matter and shear power spectra. To do so, we follow a procedure similar to the one of Ref. \citep{Audren+12}.  In that work, the authors presented a forecast of the errors on cosmological parameters expected from galaxy clustering and weak lensing observables as will be probed by a \textit{Euclid}-like survey. They performed a MCMC likelihood analysis assuming as fiducial, ``mock'' data the theoretical prediction (i.e. without statistical scatter). This had been shown to lead to the same forecasted errors as employing more realistic measurements from N-body simulations \cite{Perotto+06}. In particular, the modelling of the non-linear power spectrum uses the \textsc{Halofit} fitting functions from Ref.~\cite{Bird-HALOFIT+12}, available at the time and the fiducial cosmic shear spectra were computed using \eq{eq:shear}.  The systematic uncertainty affecting the \textsc{Halofit} prescription was accounted for in the error budget.

In the following subsections we describe in detail the similar analysis we perform for the clustering and shear survey along with the specific characteristics assumed for the surveys, borrowed as well from Ref. \cite{Audren+12}.

\subsection{Clustering survey}

As already mentioned, we are limiting the scope of our analysis to exploring degeneracies at the level of the matter density field. As our reference, 3-D matter distribution we consider the volume probed by a spectroscopic redshift survey covering a sky fraction of $f_\textrm{sky} = 0.375$, spanning a redshift range from   $z = 0.5$ to $z= 2.0$ subdivided in 16 redshift bins. The volume of each bin is given by
\be
V_\mathrm s(z) = 4\pi \ f_\textrm{sky} \ \chi^2(z) \ \frac{\Delta z}{(1+z)^3} \ \der{\chi}{z},
\label{eq:volume_survey}
\ee
where $\Delta z = 0.1$ is the bin width in redshift. Since our ``observable'' is the three-dimensional matter power spectrum, additional, relevant survey characteristic such as the expected galaxy number density will not play a role in our analysis.

We assume a Gaussian likelihood function for cold matter power spectrum $P_\mathrm{cc}$ given by
\be
\ln \mathcal{L}_C \propto -\frac{1}{2} \sum_{z \ \mathrm{bins}} \sum_{i,j} \left[P^\mathrm{th}_\textrm{cc}(k_i,z) - P^\mathrm{obs}_\textrm{cc}(k_i,z)\right]
\left[\mathcal{C}(z)^{-1}\right]_{ij}
\left[P^\mathrm{th}_\textrm{cc}(k_j,z) - P^\mathrm{obs}_\textrm{cc}(k_j,z)\right],
\label{eq:likelihood_Pk}
\ee
where  $P^\mathrm{th}_\textrm{cc}(k,z)$ and $P^\mathrm{obs}_\textrm{cc}(k,z)$} are respectively the theoretical and ``observed'' cold matter power spectra while $\mathcal{C}_{ij}$ is the covariance matrix. All power spectra are evaluated in wavenumber bins of size $\Delta k=0.0163\kMpc$ from a minimal value of $k_\textrm{min}=0.01\kMpc$ to a maximum of $k_\textrm{max}=0.5\kMpc$. The chosen value of $\Delta k$ is always larger  than the effective fundamental frequency $k_\textrm f^\textrm{eff}(z)\equiv 2\pi/[V_\textrm{s}(z)]^{1/3}$ of each subvolume $V_\textrm{s}(z)$ defined by the binning in redshift. The value for $k_\textrm{max}$ is a rather optimistic estimate for the maximum scale that future surveys will reach. In fact, most spectroscopic surveys targeting baryonic oscillations as one of the main cosmological probes are, by design, limited by shot-noise to $k_\textrm{max}=0.3-0.4 \kMpc$.

All power spectra are generated using the Code for Anisotropies in the Microwave Background (\camb) \cite{Camb}. Non-linearities are modelled through the version of the \textsc{Halofit} module \cite{Mead-b+15, Mead-nu+16}.  Since we are interested in the cold matter power spectrum $P_\textrm{cc}$, however, we cannot use the default output of the \halofit module in \camb  out-of-the-box because that only applies to the total matter power spectrum. Rather we apply \halofit as a nonlinear operator, denoted below as  $\mathcal{H}$, on the linear cold matter power spectrum. The effects of massive neutrinos are therefore accounted for mainly via the linear power spectrum $P_\textrm{cc}^L$, with an additional, mild dependence on the neutrino fraction $f_\nu$ in the \halofit parameters $\delta_\mathrm c$ and $\Delta_v$ described in eq.\,(25) of Ref. \cite{Mead-nu+16}. 

In addition, neutrino effects on matter power spectrum have been shown to be separable from the baryon feedback ones \cite{Mummery-fb_vs_nu+17}, so we implement baryonic effects by means of \eq{eq:feedback_function} as a multiplicative factor $F_{\rm bf}$ to the non-linear cold matter power spectrum alone. It follows that the \halofit module parameters providing the nonlinear mapping $P_\mathrm{cc}=\mathcal{H}(P^L_\mathrm{cc})$ will correspond to the ``DM only'' case (see Table 4 in Ref. \cite{Mead-b+15}) as we are treating the baryonic suppression separately. In short, our model for the nonlinear cold matter power spectrum will be given by 
\be
P^\textrm{obs/th}_\mathrm{cc}(k,z) = \mathcal{H}\left[P_\mathrm{cc}^\mathrm{L}(k,z)\right] F_\mathrm{bf}(k,z|M_c, \eta_\mathrm b, z_\mathrm c)\,.
\label{eq:power_spectrum_observed}
\ee
The accuracy of these fitting formulae will be taken into account in the error budget as we will discuss shortly.

In fact, the covariance matrix $\mathcal C_{ij}$ in \eq{eq:likelihood_Pk} accounts for both statistical and systematic errors, as 
\be
\mathcal{C}_{ij}(z) =
\mathcal{C}^\mathrm{CV}_{ij}(z) + \mathcal{C}^\mathrm{sys}_{ij}(z) \,,
\ee
where the statistical contribution from cosmic variance is simply given by
\be
\mathcal{C}^\mathrm{CV}_{ij}(z) =
\frac{(2\pi)^2}{k_i^2 \ \Delta k \  V_\mathrm s(z)}  \ P_\textrm{cc}^2(k_i,z)\ \delta_{ij} \,,
\ee
and therefore neglects any non-Gaussian component along with any beat-coupling/super-sample covariance term from the finite observed volume
\cite{Bispectrum-Sefusatti+06, Covariance-Hamilton+05, Super_sample_covariance-Takada+13}.

Concerning instead the systematic uncertainty affecting the theoretical predictions for the matter power spectra, our standpoint assumes that such predictions are the outcome of state-of-art numerical simulations (we do {\em not} consider here the alternative, analytical approach based on perturbation theory). As such they will suffer from systematic uncertainties related to the choice of the N-body code, the resolution, etc. but also on the accuracy of the fitting function or the emulator built to exploit the numerical results in an efficient MC-based likelihood analysis of future data-sets.     

A recent code comparison \cite{Precision_Pk-Schneider+16} estimates the systematic difference among different codes at the 1\% and 3\% level respectively at $k=1$ and $10\kMpc$, while suggesting a minimum box size and maximum particle mass in order to avoid errors due to finite-volume effects and resolution beyond the percent level. Similar comparisons are not available for numerical simulations of massive neutrino cosmologies although one can expect additional errors of the order of 1\% related to the problem of the proper setting of initial conditions \cite{Zennaro+17}. Clearly, we do not include in this budget the error on the numerical description of baryonic effects since we deal with it explicitly: the evaluation of its impact is precisely the goal of this work.  

In addition to the systematic error intrinsic to the numerical approach, fitting functions such as the various versions of \halofit \cite{Halofit-Smith+03,Takahashi+12,Bird-HALOFIT+12} are also affected by their own uncertainty in reproducing the N-body results on which they are based.
Here, for instance, we use the newest version of \halofit, based on the \hmcode \cite{Mead-nu+16}.
The latter
is expected to provide an accuracy of a few percent at $k<10\kMpc$ for the most common extensions to the Standard Model only worsening to 10\% for specific modified gravity models with chamelon screening. On the other hand, another viable approach is given by cosmic emulators.  The \cosmicemu code \cite{CoyoteI, CoyoteII, CoyoteIII, Cosmic_emu+10}, in its latest version, is claimed to predict the power spectrum at the 1\% level up to $k=8\kMpc$ over a significant region of the allowed parameter space.
The accuracy of the new \textsc{EuclidEmulator} \cite{EuclidEmulator} is approximately $\sim 1 \%$  at $k< 1 \kMpc$, therefore comparable to the one obtained with N-body simulations.
Finally, an alternative method, combining perturbation theory results at large scales and fitting functions from high-resolution simulations at small scales, has recently been proposed by Ref. \cite{Smith-Angulo+18} and implemented in the \NGhalofit code, characterised by an expected accuracy of 1\% up to $k=0.9\kMpc$.

In light of these results, we will take a conservative stand assuming that N-body-based methods providing predictions for the nonlinear power spectrum, including massive neutrino effects and common extensions to the Standard Model, are affected by a systematic uncertainty of 3\% at $k=1\kMpc$ and 4\% at $k=10\kMpc$ for our Planck fiducial cosmology at $z=0$. In particular, following \cite{Bird-HALOFIT+12}, we describe the scale and redshift dependence of the relative, systematic error on the cold matter power spectrum as
\be
\alpha(k,z) \equiv \frac{\Delta P_\mathrm{cc}(k,z)}{P_\mathrm{cc}(k,z)} = \frac{\ln[1+k/k_\sigma(z)]}{1+\ln[1+k/k_\sigma(z)]} \ 5\%,
\label{eq:halofit_error}
\ee
where $k_{\sigma}(z)$ is the non-linear scale at which the mass fluctuations, smoothed by a Gaussian filter, are equal to unity, that is $\sigma_R=1$ for $R=1/k_\sigma$. This functional form for $\alpha(k,z)$ was also used by Ref. \cite{Audren+12}, while a different form for the same quantity has been adopted by Ref. \cite{Sprenger+18}.
At the value of $k_\textrm{max}=0.5\kMpc$ adopted for the clustering analysis we have $\alpha\simeq 2\%$ at redshift zero,  which is a conservative choice.  

We expect any systematic error to correlate different wavenumbers. Following Ref. \cite{Baldauf+16}, we account for the systematic uncertainty with a contribution to the covariance matrix $C_{ij}$ of the form  
\be
\mathcal{C}^\mathrm{sys}_{ij}(z) =
\alpha(k_i,z)\,  \alpha(k_j,z) \ P_\textrm{cc}(k_i,z) \ P_\textrm{cc}(k_j,z) \ 
\exp{\left[-\frac{\log^2(k_i/k_j)}{2\sigma_\alpha^2}\right]},
\label{eq:covmat_Pk}
\ee
where the log-exponential term represents the correlation kernel. We choose $\sigma_\alpha = \log 5$ as the log-scale correlation length.
This choice is motivated by the fact that the typical scale over which \textsc{Halofit} overestimates/underestimates the power spectrum of a simulation is roughly half a decade (see for example Figure 1 of Ref. \cite{Mead-b+15}).

In this work we use the \textsc{Emcee}\footnote{\url{http://dfm.io/emcee/current/}} package for the likelihood sampling.

\subsection{Cosmic shear survey}

For the cosmic shear survey we assume as well a sky coverage of $f_{\rm sky}=0.375$.
The distribution of source galaxies is taken to be
\be
n_\mathrm s(z) = \frac{\beta}{z_0 \ \Gamma\left(\frac{\alpha+1}{\beta}\right)} \ \left(\frac{z}{z_0}\right)^\alpha \ \exp{\left[-\left(\frac{z}{z_0}\right)^\beta\right]},
\ee
where the constants in front ensure that the integral over all redshifts of $n_\mathrm s(z)$ is normalized to 1.
We set $\alpha = 2$, $\beta = 1.5$ and $z_\mathrm{mean} = 1.412 \ z_0 = 0.9$.
We fit a multipole range going from $\ell=10$ up to $\ell = 2000$, corresponding to an angle of 5 arcseconds or, equivalently, a scale of $\sim 0.7 \kMpc$ at the median redshift.
We divide the sources into $N=3$ redshift bins, in such a way that each of them contains the same number of galaxies.
We assume a galaxy density of $d = 30 \ \unit{arcmin^{-2}}$ with a root mean square intrinsic shear of $\mean{\gamma_\mathrm{rms}^2}^{1/2}=0.30$.
All the values and functional forms employed here are likely to be in broad agreement  with (or could easily be generalized to) future surveys specifics like \textit{Euclid}.
In terms of these quantities we can estimate the shape noise to be given by
\be
N^{(ij)}(\ell) = \delta_{ij} \ \mean{\gamma_\mathrm{rms}^2} \ n_i^{-1},
\label{eq:shear_noise}
\ee
where $n_i = 3600 \ d \left(180/\pi\right)^2/N$ is the number of galaxies per steradian in the $i$-th bin.

The likelihood function is Gaussian also for the shear power spectrum $C_{\ell}^{(ij)}$ and is given by
\bea
\ln \mathcal{L}_S &\propto& -\frac{1}{2} \sum_{i,j} \sum_{\ell, \ell'}
\left(C_\mathrm{obs}^{(ij)} - C_\mathrm{th}^{(ij)} \right)_{\ell} \left[\mathcal{C}^{(ij)}\right]^{-1}_{\ell\ell'}
\left(C_\mathrm{obs}^{(ij)} - C_\mathrm{th}^{(ij)} \right)_{\ell'},
\label{eq:likelihood_Cl}
\eea
where $\mathcal{C}_{\ell\ell'}^{(ij)}$ represents the power spectra covariance matrix with the indices $i,j$ running from 1 to 3 labelling the redshift bins and $\ell, \ell'$ labelling the multipoles.

Consistently with the discussion in the previous Section,  the total matter power spectrum, relevant for weak lensing observables, is obtained as
\bea
P_\mathrm{mm}(k,z) & = & (1-f_\nu)^2 \ \mathcal{H}\left[P_\mathrm{cc}^\mathrm{L}(k,z)\right] \ F_\mathrm{bf}(k,z|M_c, \eta_\mathrm b, z_\mathrm c) \nn\\
& & + 2 \ (1-f_\nu) \ f_\nu \ P_\mathrm{c\nu}^\mathrm{L}(k,z) + f_\nu^2 \ P_\mathrm{\nu\nu}^\mathrm{L}(k,z)\,.
\eea
Again, non-linearities are computed using the \textsc{Halofit} formulae and both the non-linear transformation and the baryonic correction apply only to the cold matter power spectrum. For the cross and neutrino spectra we use linear theory as their non-linear counterpart is expected to give sub-percent contribution \cite{DEMNUni+16}.

The sources of error we consider here are cosmic variance, shape noise and the theoretical error on the matter power spectrum propagated in the cosmic shear spectrum (\eq{eq:halofit_error_shear}), so that the total covariance matrix reads
\be
\mathcal{C}^{(ij)}_{\ell\ell'}(z) =
\mathcal{C}^{(ij), \mathrm{CV-SN}}_{\ell\ell'}(z) + \mathcal{C}^{(ij), \mathrm{sys}}_{\ell\ell'}(z)\, .
\label{eq:covmat_Cl}
\ee
The cosmic variance, in the Gaussian approximation, and shape noise contributions is given by
\be
\mathcal{C}^{(ij), \mathrm{CV-SN}}_{\ell\ell'}=
\frac{2}{2\ell+1} \ f_{\mathrm{sky}}^{-1} \ \left. \left[C_\ell^{(ij)} + N_\ell^{(ij)} \right]\right.^2 \ \delta_{\ell\ell'} \,.
\ee
The systematic component is given instead by
\be
 \mathcal{C}^{(ij), \mathrm{sys}}_{\ell\ell'} =
E_{\ell}^{(ij)} \ E_{\ell'}^{(ij)} \exp{\left[-\frac{\log^2(\ell/\ell')}{2 \ \sigma_\mathrm E^2}\right]}.
\ee
where the relative uncertainty on the shear power spectrum is obtained by propagating the uncertainty on the matter power spectrum through \eq{eq:shear} as
\bea
E^{(ij)}(\ell) & \equiv & \frac{\Delta C^{(ij)}(\ell)}{C^{(ij)}(\ell)} = \nn\\
& = &  \int_0^\infty \!\!\!\de z \ \frac{c}{H(z)} \ \frac{W^{(i)}(z) \ W^{(j)}(z)}{\chi^2(z)} \ \alpha\left(k=\frac{\ell}{\chi(z)},z\right) \ P_\mathrm{mm}\left(k=\frac{\ell}{\chi(z)}, z \right)\,.
\label{eq:halofit_error_shear}
\eea
Here we have implicitly assumed that the error on $P_\mathrm{mm}$ is the same of that on $P_\mathrm{cc}$, since the other quantities are involved at the linear level and are therefore known with high precision. 

The value for the correlation length $\sigma_\mathrm E$ for the error on the shear power spectrum is chosen consistently with the one on the matter power spectrum. To estimate it we introduce a logarithmic modulation of period $\sigma_\alpha$ in the matter power spectra and computed the shear spectra integrating them.
This was translated in a modulation of period approximately one third of a decade in the shear spectra: hence we set $\sigma_\mathrm E = \log 3$.

As a final remark, we should stress that, despite a mission like \textit{Euclid} will measure clustering and shear in the same patch of the sky, we never perform a combined analysis of the two quantities. We will leave this for future work.

\section{Results}
\label{sec:results}

This Section presents our results and is divided into three parts.
In the first, we verify if a properly chosen set of feedback parameters is able to reproduce a suppression in the matter spectra similar to the effect of  massive neutrinos. In the second, we address the possible degeneracies between the neutrino mass and the three feedback parameters. In the third, we investigate additional, possible degeneracies between $M_\nu$ and the intrinsic alignment parameter as it could be measured in weak lensing surveys.

\subsection{Fitting baryon feedback on massive neutrino cosmologies}
\label{sec:feedback_on_nu}

The goal of this Section is to check whether there exists a set of reasonable feedback parameters which is able to reproduce the same effects of massive neutrinos.
We adopt as fiducial cosmology a model with a single massive neutrino species and we assume no baryon feedback.
We consider a single massive neutrino species with mass $M_\nu = 0.15, 0.30, 0.45$ eV. 
We look for a fit to such mock data with a model that assumes a constant neutrino mass corresponding to the minimum allowed value $M_\nu = 0.056$ eV \cite{LP_massive_nu+06} but with varying baryon feedback parameters. 

\begin{table}[t]
\centering
\renewcommand{\arraystretch}{1.3}
\scalebox{0.94}{
\begin{tabular}{c||c|c|c||c||c|c|c||c}
\hline
& \multicolumn{4}{c||}{ \textbf{Matter clustering}, $P(k)$} & \multicolumn{4}{c}{ \textbf{Cosmic shear}, $C_\ell$} \\
\hline
 $M_\nu$ &  $\logMc$ &  $\etab$ & \Large $z_c$ & $\Delta \chi_\mathrm{red}^2$ &  $\logMc$ & $\etab$ & $z_c$ & $\Delta \chi_\mathrm{red}^2$\\
\hline
\hline
 0.15 eV &
$12.56^{+0.02}_{-0.02}$ & $5.5^{+2.3}_{-0.8}$ & $>8.3$ & \textbf{+0.03}
& $12.77^{+0.11}_{-0.11}$ & $14.0^{+10.8}_{-8.9}$ & $>4.8$ & \textbf{+0.005}\\
\hline
 0.30 eV &
$13.33^{+0.02}_{-0.02}$ & $3.6^{+0.1}_{-0.1}$ & $>10.0$ & \textbf{+0.27}
& $13.85^{+0.11}_{-0.11}$ & $3.3^{+2.9}_{-0.9}$ & $>5.2$ & \textbf{+0.009}\\
\hline
 0.45 eV &
$14.08^{+0.01}_{-0.01}$ & $2.6^{+0.1}_{-0.1}$ & $>12.3$ & \textbf{+0.95}
& $14.86^{+0.09}_{-0.09}$ & $1.8^{+0.4}_{-0.3}$ & $>5.1$ & \textbf{+0.019}\\
\hline
\end{tabular}
}
\caption{
Best-fit values of the baryon parameters obtained from the analysis of Section \ref{sec:feedback_on_nu}, where we fitted spectra with baryonic features onto spectra containing massive neutrinos. We also report the difference in the reduced chi-squared $\Delta\chi_\mathrm{red}^2$ with respect to the one obtained using the ``true'' model. $M_c$ is in units of $M_\odot/h$ while the errors or lower limits represent the 68\% confidence level. The priors are $\log{M_c} \ [M_\odot/h] \in [12,30]$, $\eta_\mathrm b \in [0,30]$, $z_\mathrm c \in [0,30]$.
}
\label{tab:3p}
\end{table}

\begin{figure}[!t]
	 \makebox[1\textwidth][c]
		{
			\includegraphics[width=0.65\textwidth]{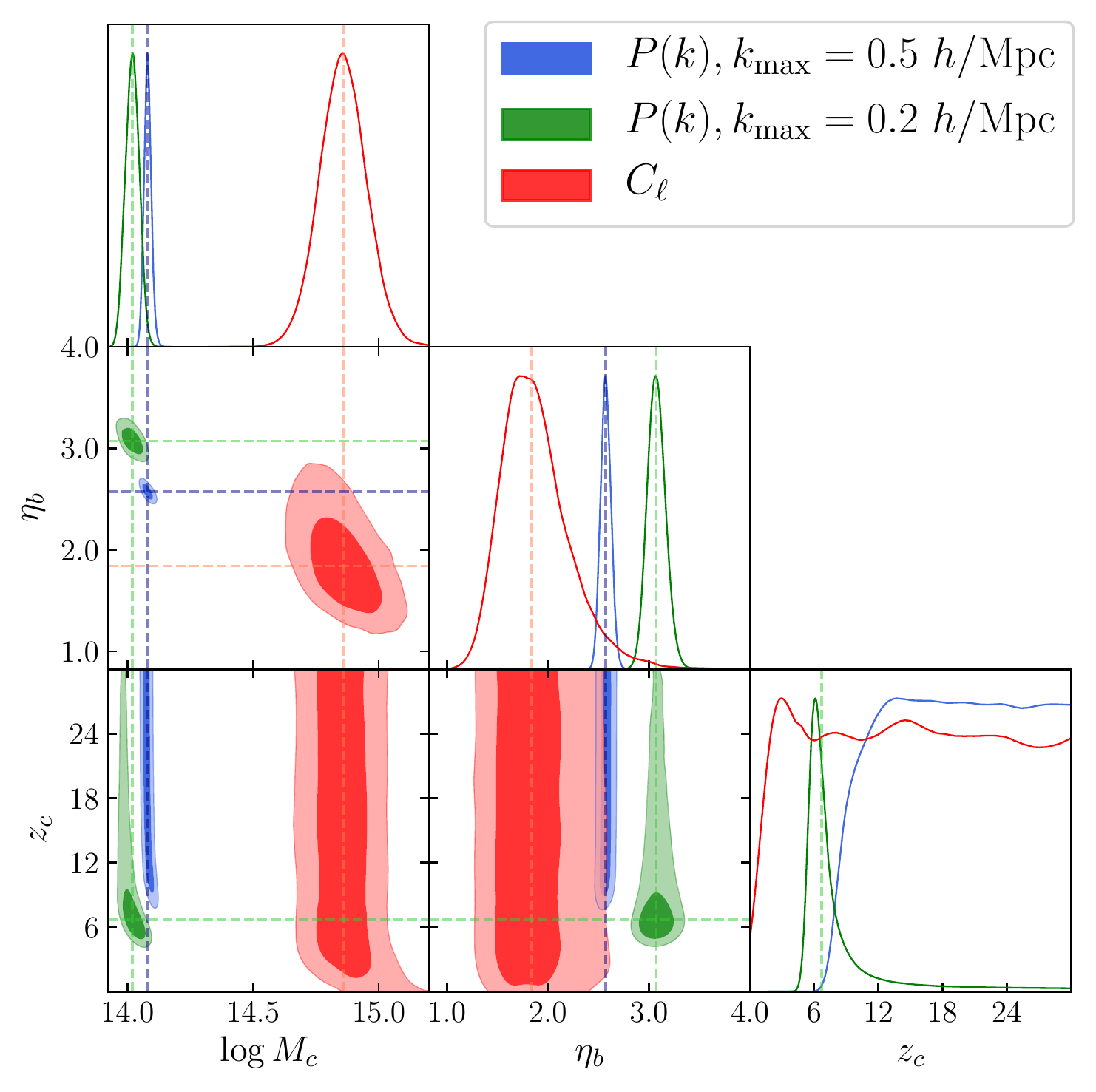}
   		  		 } 
\caption{Contour plots showing the posterior probability distribution for the three feedback parameters. These are obtained when matter (blue and green, where we stop at 2 different $k_\mathrm{max}$) and shear (red) spectra for cosmologies with minimal neutrino mass ($M_\nu = 0.056$ eV) and baryon feedback are fitted on spectra in massive neutrino cosmologies ($M_\nu = 0.45$ eV in this plot) with no baryon feedback.}
\label{fig:contour_m0450_3p}
\end{figure}

The results of this test are summarised in Table \ref{tab:3p}, which reports the best-fit values for the parameters, along with their 68\% confidence level errors. The table also shows the shift in the reduced chi-square with respect to the one obtained by fitting the true model on the mock data.
The first thing to notice is that clustering can better distinguish between the two effects - massive neutrinos and baryon feedback -  even for the lowest neutrino masses.
This can be addressed to the larger cosmic variance that one has in weak lensing surveys, that dominates the total error at almost all multipoles (see for example Figure \ref{fig:Cl_nu_feed}). 
Furthermore, the characteristic scales are more clearly defined for the three-dimensional power spectrum than for the shear one.

Interestingly, we find some discrepancies in the contour plots in Figure \ref{fig:contour_m0450_3p}: it seems not possible to find feedback parameters able to describe the neutrino-induced suppression in both the observables. In fact, the picture shows the allowed (1- and 2-$\sigma$) regions for the feedback parameters inferred by cosmic shear (red) and clustering when stopping the analysis at $k_\mathrm{max} = 0.5 \kMpc$ (blue) or at $k_\mathrm{max} = 0.2 \kMpc$ (green), for the case $M_\nu = 0.45$ eV.
Not only the clustering contours do not match their counterpart from cosmic shear, but there is a significant dependence on the maximum wavenumber, $k_\mathrm{max}$ assumed for the analysis. Despite the three degrees of freedom describing the baryonic effects, the fit is even more inaccurate at the intermediate scales between $0.2$ and $0.5 \kMpc$: for instance, when stopping at $k = 0.2 \kMpc$ for $M_\nu = 0.45$ eV we obtained $\Delta\chi^2_\mathrm{red} = 0.73$, versus a $\Delta\chi^2_\mathrm{red} = 0.95$ when pushing up to $k=0.5 \kMpc$.
This provides some first clue that a combined analysis should be able to disentangle baryonic and neutrinos effects. However, since we expect a high degree of correlation between the two observables, a detailed and careful study of their covariance matrices must be carried out.

The redshift parameter $z_\mathrm c$ remains largely unconstrained whatever the neutrino mass. The bottom right panel of Figure \ref{fig:Pk_nu_feed} gives a simple explanation for that. At $z = 1$ (a characteristic redshift for future galaxy surveys) and at scales of $0.5\kMpc$ (that is the maximum we consider for our analysis) the dependence of the power spectrum on $z_c$ is very weak: the difference with respect to the ``base'' model lies within the theoretical uncertainty. Besides, the suppression almost saturates for $z_\mathrm c \gtrsim 3$, meaning that any value for this parameter is equivalent. The same holds true also for the shear power spectrum (see the bottom right panel of figure \ref{fig:Cl_nu_feed}): $z_\mathrm c$ does not play an important role, as the variation in the range $z_\mathrm c = 1-3$ is of the order of 2\% versus a theoretical uncertainty of $\sim 3\%$ even at the highest multipoles.

The $\eta_\mathrm b$ parameter sets the scale at which the suppression occurs, so we expect that it is somewhat degenerate with neutrino mass.
In fact Table \ref{tab:3p} shows this effect: a higher neutrino mass implies a lower value for $\eta_\mathrm b$.
Moreover, with increasing neutrino mass the constraints on this parameter get tighter.
The reason is two-fold: firstly, because of the strong power law dependence of the feedback suppression on $\eta_\mathrm b$ (see \eq{eq:feedback_function}); secondly, a small neutrino mass implies a lower suppression at large scales, where cosmic variance is larger.

\subsection{Baryonic feedback and massive neutrinos degeneracies}
\label{sec:feedback_nu}

We address now directly the degeneracy between the neutrino mass and the parameters of the BCM by considering a likelihood analysis where all relevant parameters are allowed to vary simultaneously.

We consider 16 distinct fiducial models corresponding to all possible combinations for each of the four parameters taking the two values given by  $M_\nu = (0.15, 0.3)$ eV, $\logMc \ [M_\odot/h]= (13, 14)$, $\etab = (0.5, 1)$, $z_\mathrm c = (1, 2)$.
We then run the MCMC likelihood analysis over the four parameters with a twofold goal: check whether we are able to recover the fiducial values, with special attention to the neutrino mass, and examine the degeneracies among the parameters.

\begin{table}[!t]
\centering
  \renewcommand{\arraystretch}{1.5}
  \resizebox{.9\columnwidth}{!}{
\begin{tabular}{cccc|ccc|ccc}
\multicolumn{4}{c|}{ Input parameters}  & \multicolumn{3}{c|}{ $P(k)$} & \multicolumn{3}{c}{ $C_\ell$} \\
  \toprule
 $M_\nu$ & \!\!$\logMc $\!\! & $\etab$ & $z_\mathrm c$ &  $M_\nu$ &  $\logMc$ & $\etab$ & $M_\nu$ &  $\logMc$ & $\etab$\\
  \hline
  \midrule
0.15 &  13 &  0.5 & 1.0 &
  $\mathbf{0.150^{+0.006}_{-0.009}}$ & $<14.6$ & \xmark &
  $\mathbf{0.147^{+0.030}_{-0.045}}$ & $<13.3$ & $<5.6$ \tabularnewline
0.15 &  13 &  0.5 & 2.0 &
  $\mathbf{0.150^{+0.004}_{-0.004}}$ & \xmark & \xmark &
  $\mathbf{0.148^{+0.033}_{-0.041}}$ & $<13.4$ & $<8.3$ \tabularnewline
0.15 &  13 &  1.0 & 1.0 &
  $\mathbf{0.150^{+0.003}_{-0.003}}$ & $13.2^{+0.6}_{-0.3}$ & $0.9^{+0.2}_{-0.3}$ &
  $\mathbf{0.167^{+0.048}_{-0.058}}$ & $<13.2$ & $<5.5$ \tabularnewline
0.15 &  13 &  1.0 & 2.0 &
  $\mathbf{0.150^{+0.003}_{-0.003}}$ & $13.0^{+0.1}_{-0.1}$ & $1.0^{+0.1}_{-0.1}$ &
  $\mathbf{0.156^{+0.050}_{-0.054}}$ & $<13.3$ & $<2.3$ \tabularnewline
  
0.15 &  14 &  0.5 & 1.0 &
  $\mathbf{0.150^{+0.003}_{-0.003}}$ & $<16.9$ & $<1.0$ &
  $\mathbf{0.152^{+0.034}_{-0.037}}$ & $13.9^{+0.3}_{-0.3}$ & $0.5^{+0.1}_{-0.1}$ \tabularnewline
0.15 &  14 &  0.5 & 2.0 &
  $\mathbf{0.150^{+0.003}_{-0.003}}$ & $14.4^{+1.9}_{-1.1}$ & $0.4^{+0.3}_{-0.2}$ &
  $\mathbf{0.150^{+0.034}_{-0.037}}$ & $14.0^{+0.3}_{-0.3}$ & $0.5^{+0.1}_{-0.1}$ \tabularnewline
0.15 &  14 &  1.0 & 1.0 &
  $\mathbf{0.150^{+0.003}_{-0.003}}$ & $14.0^{+0.1}_{-0.1}$ & $1.0^{+0.1}_{-0.1}$ &
  $\mathbf{0.166^{+0.061}_{-0.059}}$ & $13.8^{+0.4}_{-0.4}$ & $1.0^{+0.2}_{-0.2}$ \tabularnewline
0.15 &  14 &  1.0 & 2.0 &
  $\mathbf{0.150^{+0.003}_{-0.003}}$ & $14.0^{+0.1}_{-0.1}$ & $1.0^{+0.1}_{-0.1}$ &
  $\mathbf{0.165^{+0.062}_{-0.062}}$ & $13.9^{+0.4}_{-0.5}$ & $1.0^{+0.2}_{-0.2}$ \tabularnewline

\hline
0.30 &  13 &  0.5 & 1.0 &
  $\mathbf{0.300^{+0.005}_{-0.005}}$ & $<16.4$ & \xmark &
  $\mathbf{0.300^{+0.031}_{-0.051}}$ & $<13.5$ & $<8.3$ \tabularnewline
0.30 &  13 &  0.5 & 2.0 &
  $\mathbf{0.300^{+0.007}_{-0.004}}$ & $<18.3$ & \xmark &
  $\mathbf{0.303^{+0.033}_{-0.055}}$ & $<13.4$ & $<11.3$ \tabularnewline
0.30 &  13 &  1.0 & 1.0 &
  $\mathbf{0.300^{+0.003}_{-0.003}}$ & $13.1^{+0.6}_{-0.3}$ & $0.9^{+0.2}_{-0.3}$ &
  $\mathbf{0.310^{+0.054}_{-0.080}}$ & $<13.4$ & $<6.9$ \tabularnewline
0.30 &  13 &  1.0 & 2.0 &
  $\mathbf{0.300^{+0.003}_{-0.003}}$ & $13.0^{+0.1}_{-0.1}$ & $1.0^{+0.1}_{-0.1}$ &
  $\mathbf{0.317^{+0.051}_{-0.084}}$ & $<13.7$ & $<5.1$ \tabularnewline
  
0.30 &  14 &  0.5 & 1.0 &
  $\mathbf{0.300^{+0.003}_{-0.003}}$ & $<16.6$ & $<0.9$ &
  $\mathbf{0.301^{+0.035}_{-0.040}}$ & $13.9^{+0.3}_{-0.3}$ & $0.5^{+0.1}_{-0.1}$ \tabularnewline
0.30 &  14 &  0.5 & 2.0 &
  $\mathbf{0.300^{+0.003}_{-0.003}}$ & $14.4^{+1.7}_{-1.1}$ & $0.4^{+0.3}_{-0.2}$ &
  $\mathbf{0.299^{+0.036}_{-0.041}}$ & $14.0^{+0.3}_{-0.3}$ & $0.5^{+0.1}_{-0.1}$ \tabularnewline
0.30 &  14 &  1.0 & 1.0 &
  $\mathbf{0.300^{+0.003}_{-0.003}}$ & $14.0^{+0.1}_{-0.1}$ & $1.0^{+0.1}_{-0.1}$ &
  $\mathbf{0.322^{+0.068}_{-0.086}}$ & $13.7^{+0.6}_{-0.5}$ & $1.0^{+0.2}_{-0.2}$ \tabularnewline
0.30 &  14 &  1.0 & 1.0 &
  $\mathbf{0.300^{+0.003}_{-0.003}}$ & $14.0^{+0.1}_{-0.1}$ & $1.0^{+0.1}_{-0.1}$ &
  $\mathbf{0.318^{+0.070}_{-0.085}}$ & $13.9^{+0.6}_{-0.5}$ & $1.0^{+0.2}_{-0.2}$ \tabularnewline
  \bottomrule
  \end{tabular}
 }
\caption{Best-fit values and 68\% confidence level intervals for the parameters obtained from the power spectrum, $P(k)$, as well as from the cosmic shear, $C_\ell$, analysis. The mark \xmark \ means that such parameter is not constrained at all. See Section \ref{sec:feedback_nu} for details.}
\label{table:Pk_Cl}
 \end{table}

\begin{figure}[t]
	 \makebox[1\textwidth][c]
		{
			\includegraphics[width=.8\textwidth]{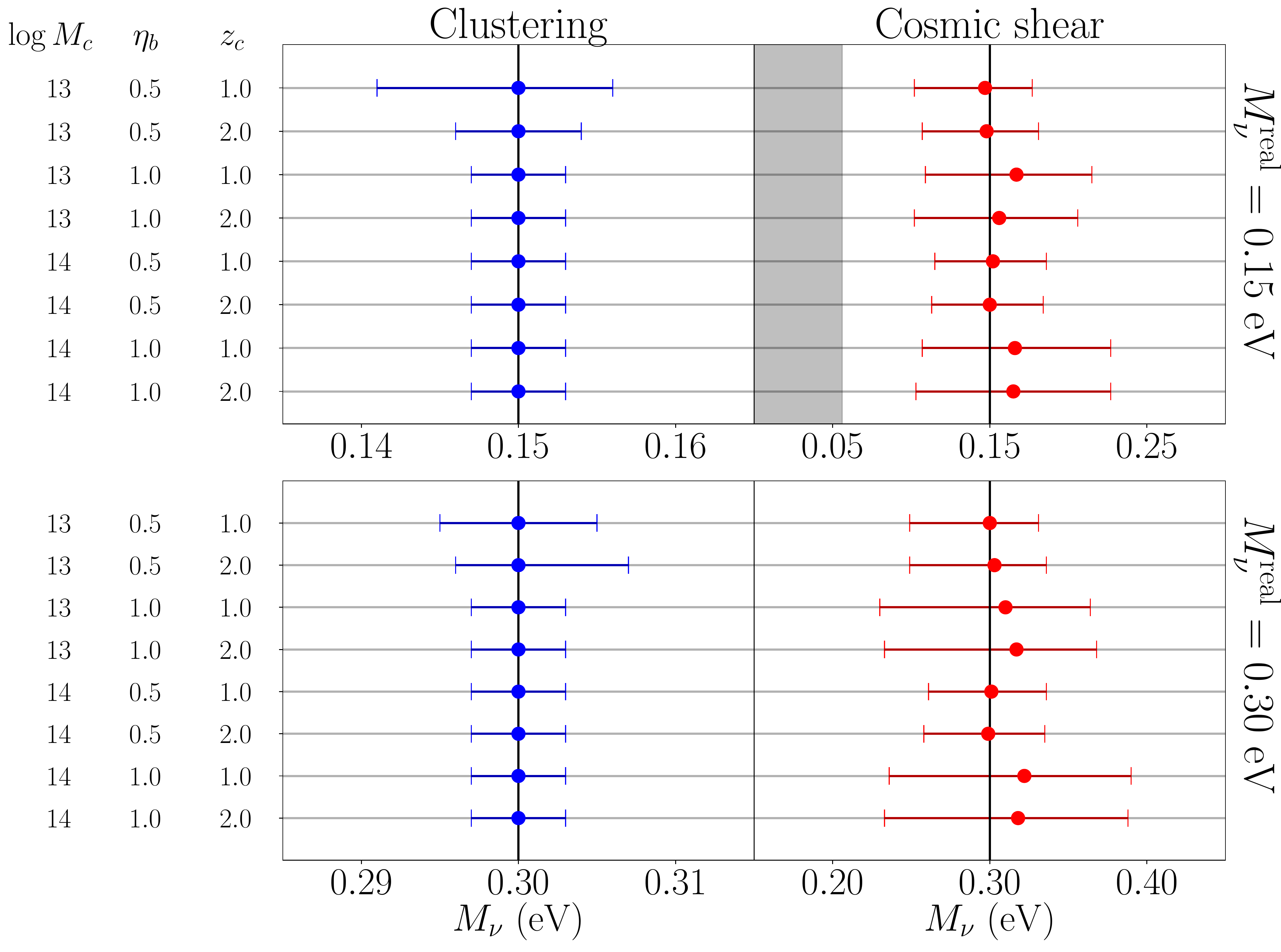}
   		  		 } 
\caption{1-D posterior probability distributions for neutrino mass for all the cases analysed in Section \ref{sec:feedback_nu}. The error-bars represent the 68\% confidence level on $M_\nu$ using clustering (blue) and weak lensing (red). The left columns show the feedback parameters used to generate the mock data. The top panels show the results when $M_\nu = 0.15$ eV, while the bottom ones do the same for the case $M_\nu = 0.30$ eV. The grey shaded area in the top panels mark the region $M_\nu < 0.056$ eV, forbidden by particle physics experiments.}
\label{fig:results_Pk_Cl}
\end{figure}

The values obtained for the parameters of major interest are listed in Table \ref{table:Pk_Cl}, while the results relative to the neutrino mass are shown in Figure \ref{fig:results_Pk_Cl}.
The blue and red data points with error bars mark the 68\% confidence level on the neutrino mass as determined, respectively, by matter clustering and cosmic shear.
We see that we can recover the right input $M^\mathrm{real}_\nu$ within 1-$\sigma$ in all cases.
The mass found with weak lensing is always within $\sim0.25\sigma$ from the right value. Moreover, the clustering ones are basically perfect (see Table \ref{table:Pk_Cl}).
The reasons for this are multiple. First of all, the matter power spectrum describes the 3-D distribution of inhomogeneities, while the shear one is a 2-D projection of a 3-D field. Thus, while the features of the matter power spectrum are well defined at each scale, the scale mixing of \eq{eq:shear} makes it difficult to associate a range of multipoles to a single effect. 
Second, neutrinos affect all multipoles in the shear power spectrum, but only the smallest scales in the matter one.
Third, the redshift dependence of the two effects is very different.
The neutrino suppression to the cdm+b power spectrum is insensitive to redshift: while the scale at which the ``turnaround'' of the spoon shape damping slightly moves towards low-$k$ values, the amplitude of such suppression stays almost constant in time.
On the other hand, Ref. \cite{Mummery-fb_vs_nu+17} shows that the suppression due to baryon feedback increases significantly at late times. Therefore tomography plays a crucial role in this kind of analysis.
Fourth, here we are assuming perfect knowledge on of the functional forms both for neutrinos and baryon feedback and that helps in recovering the correct input values with a very low level of bias.

\begin{figure}[t!]
        \centering
        \begin{subfigure}[b]{0.62\textwidth}
            \centering
            \includegraphics[width=\textwidth]{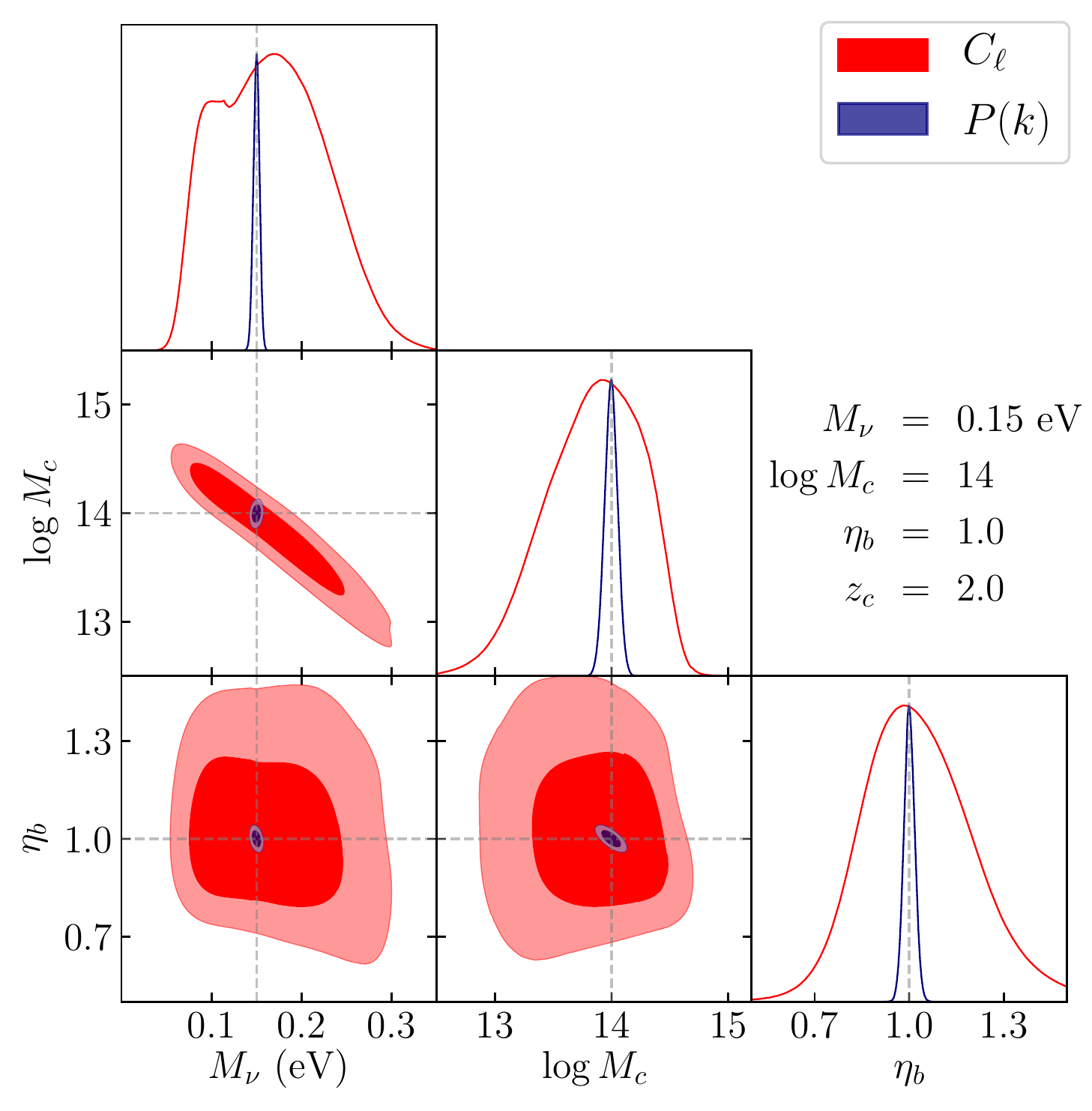}
        \end{subfigure}
        \hspace{-.7\baselineskip}
        \begin{subfigure}[b]{0.62\textwidth}  
            \centering 
            \includegraphics[width=\textwidth]{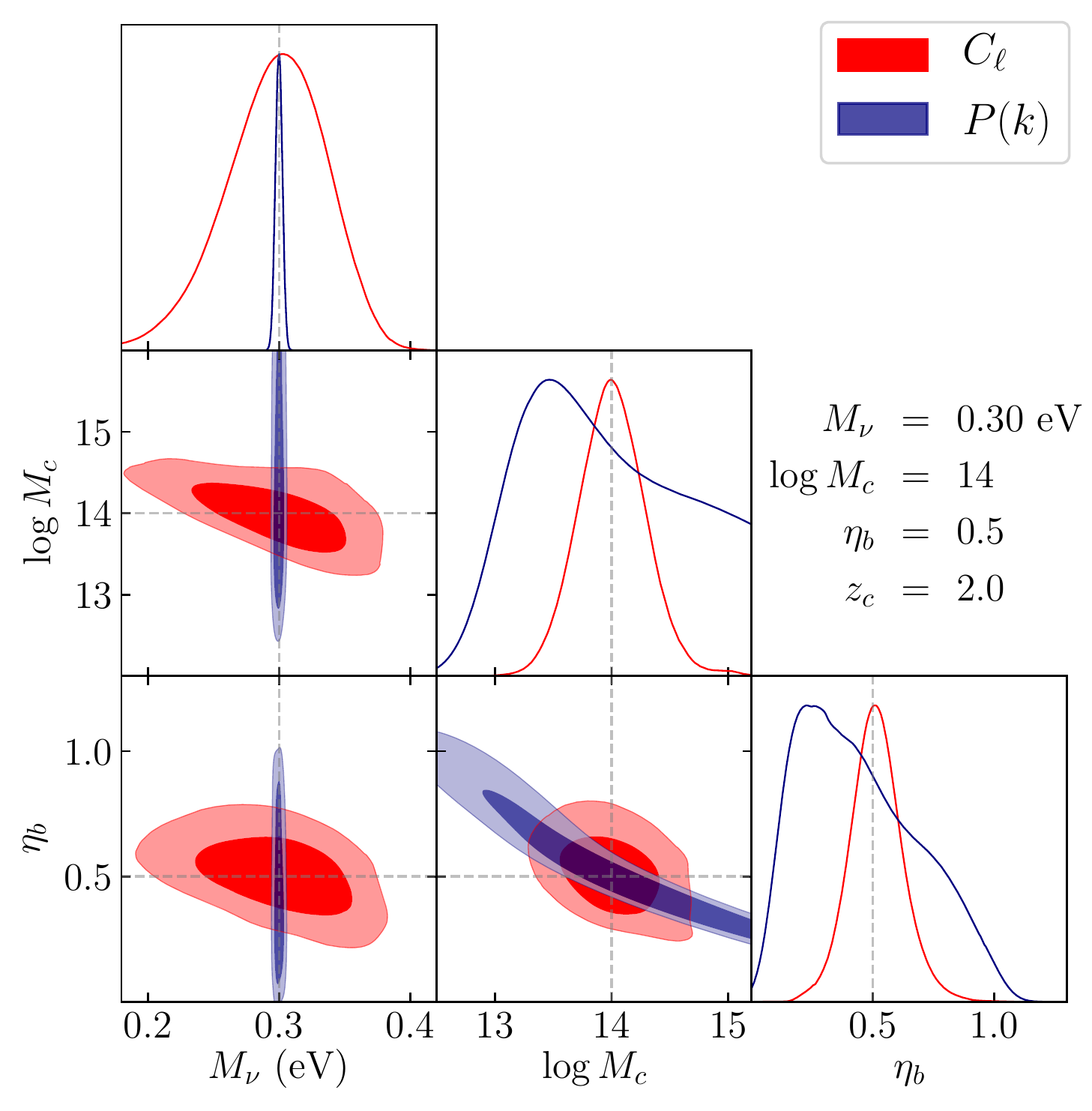}
        \end{subfigure}
\caption{1-D and 2-D posterior PDFs for $M_\nu$, $\log M_\mathrm c$ and $\eta_\mathrm b$ for 2 of the 16 runs described in Sections \ref{sec:feedback_nu} whose parameters are displayed in the plot.
The blue contours show the results for clustering, while the red contours represent the results from the cosmic shear survey.
The grey dashed lines show the ``true'' values, used to generate the mock data.}
\label{fig:triangle_4p}
\end{figure}

The absence of neutrino-baryon degeneracy in the 3D clustering case is evident in Figure \ref{fig:triangle_4p}.  There we show the contour plot for 2 of the 16 runs mentioned above. The blue contours represent the 2-D posteriors on the parameters $M_\nu$, $\log M_\mathrm c$ and $\eta_\mathrm b$, marginalized over $z_\mathrm c$: the contours in the planes $M_\nu-\log M_\mathrm c$ and $M_\nu-\eta_\mathrm b$ are clearly parallel to the parameter axes, implying that neutrino mass is not degenerate with the BCM parameters.
However, there exists a degeneracy intrinsic to the feedback parameters, in particular in the $\log M_c-\eta_b$ plane, that is very clear in the bottom panel of the Figure.
On the other hand, the red contours, representing the results obtained from weak lensing, show that a degeneracy is indeed present, i.e.~the one between $M_\nu-\log M_\mathrm c$.
However, it is still weak enough not to affect the measurement on neutrino mass, which, as we mentioned above, is recovered well within the error-bars. 

Interestingly, there are some fixed triads of fiducial BCM parameters that cannot be properly constrained neither by clustering nor by weak lensing (see Table \ref{table:Pk_Cl}).
For instance, when $\log M_\mathrm c=13$ weak lensing can only return upper limits for the feedback parameters; or again low feedback activity (i.e. low $\log M_\mathrm c$ and low $\eta_\mathrm b$) is not constrained by clustering, since the scales affected by baryon feedback are mostly left out from the analysis.

The error-bars from weak lensing are typically much larger than the ones for clustering, especially those on neutrino mass.
This is expected for many reasons. First, despite performing tomography, the integration along the line-of-sight causes the loss of information on the $z$-direction.
Second, cosmic variance limits the constraining power more than in the clustering case (see Figures \ref{fig:Pk_nu_feed} and \ref{fig:Cl_nu_feed}). Moreover, for the clustering case we only considered the cdm+b power spectrum as observable: introducing galaxy bias, Alcock-Paczynski effect and RSD would considerably enlarge them.

\subsection{Degeneracy with intrinsic alignment}
\label{sec:intrinsic_alignment}

In the last part of the work we want to investigate the possible degeneracy between neutrino mass and intrinsic alignment in weak lensing measurements.
Intrinsic alignment (IA) is one of the most significant astrophysical systematics in cosmic shear surveys, as it can enhance the signal up to 10\% at all multipoles \cite{Troxel-IA+15}.
It is due to the fact that orientations of nearby galaxies can be correlated when they form and evolve in the same gravitational environment.
As a result, in a shear survey one observes not only the cosmological signal due to the large-scale gravitational lensing potential, but, in addition, an intrinsic ellipticity term, so that
\begin{equation}
\gamma_{\textrm{obs}} = \gamma_G+\gamma_I.
\end{equation}
Thus, when computing the power spectrum of this quantity, one obtains three different contributions:
\begin{equation}
C_{\textrm{obs}}^{ij}(\ell) = C_\mathrm{GG}^{ij}(\ell) + C_\mathrm{GI}^{ij}(\ell) + C_\mathrm{II}^{ij}(\ell).
\label{eq:shear_ia}
\end{equation}
The GG term is just \eq{eq:shear}, i.e. the cosmological signal.
The latter two terms represent the correlation of shapes between physically nearby galaxy pairs (II) \cite{Heavens-IA+00, Croft-IA+00} and the correlation of galaxies that are aligned with those that are lensed by the same structure (GI) \cite{Hirata-IA+04}.

To describe this systematic we use the linear alignment model of Ref. \cite{Hirata-IA+04}: large-scale correlations or fluctuations in the mean intrinsic ellipticity field of triaxial elliptical galaxies should be due to large-scale fluctuations in the primordial potential in which the galaxy formed during the matter dominated epoch.
Thus we expect a linear relation between intrinsic alignment and the matter power spectrum.
Within this paradigm, the intrinsic alignment terms read:
\begin{eqnarray}
C_\mathrm{GI}^{(ij)} (\ell)
&=&
\int_0^\infty \de z \ \frac{c}{H(z)} \ \frac{W^{(j)}(z) \ N_\mathrm s^{(i)}(z) + W^{(i)}(z) \ N_\mathrm s^{(j)}(z)}{\chi^2(z)} \times
\nonumber \\
&&
\times \ F_\mathrm{IA}\left(k = \frac{\ell}{\chi(z)},z\right) \ P_\mathrm{mm}\left(k = \frac{\ell}{\chi(z)}, z\right)
\label{eq:Cl_GI}
\\
C_\mathrm{II}^{(ij)} (\ell)
&=&
\int_0^\infty \de z \ \frac{c}{H(z)} \frac{N_\mathrm s^{(i)}(z) \ N_\mathrm s^{(j)}(z)}{\chi^2(z)} \ F_\mathrm{IA}^2\left(k = \frac{\ell}{\chi(z)},z\right) \ P_\mathrm{mm}\left(k = \frac{\ell}{\chi(z)}, z\right),
\label{eq:Cl_II}
\end{eqnarray}
with
\begin{eqnarray}
N_\mathrm s^{(i)}(z)  &=& n_\mathrm s^{(i)}(z) \ \der{z}{\chi} = n_\mathrm s^{(i)}(z) \ \frac{H(z)}{c}
\\
F_\mathrm{IA}(k,z) &=& -A_\mathrm{IA} \ C_1 \ \rho_\mathrm c \frac{\Omega_\mathrm m}{D_\mathrm{m}(k,z)},
\end{eqnarray}
where
$\rho_c$ and $\Omega_m$ are the critical density and the matter density parameter today, $D_\mathrm{m}(k,z)$ is the linear growth factor, scale-dependent for massive neutrino cosmologies,
while $C_1 = 5\times 10^{-14} h^{-2}$ $\unit{M_\odot^{-1} \ Mpc^3}$ is a normalization constant chosen such that the intrinsic alignment free parameter $A_\mathrm{IA}$ takes values around unity.
For instance, Ref. \cite{Koh-KiDS+17} found $A_\mathrm{IA} = -1.81^{+1.61}_{-1.21}$ and $A_\mathrm{IA} = -1.72^{+1.49}_{-1.25}$ for the  analyses using 3$-z$ and 2$-z$ bins respectively, while Ref. \cite{DES_lensing+17}, although using another model, obtained $A_\mathrm{IA}= 1.3^{+0.5}_{-0.6}$.

\begin{figure}[!t]
	 \makebox[1\textwidth][c]
		{
			\includegraphics[width=\textwidth]{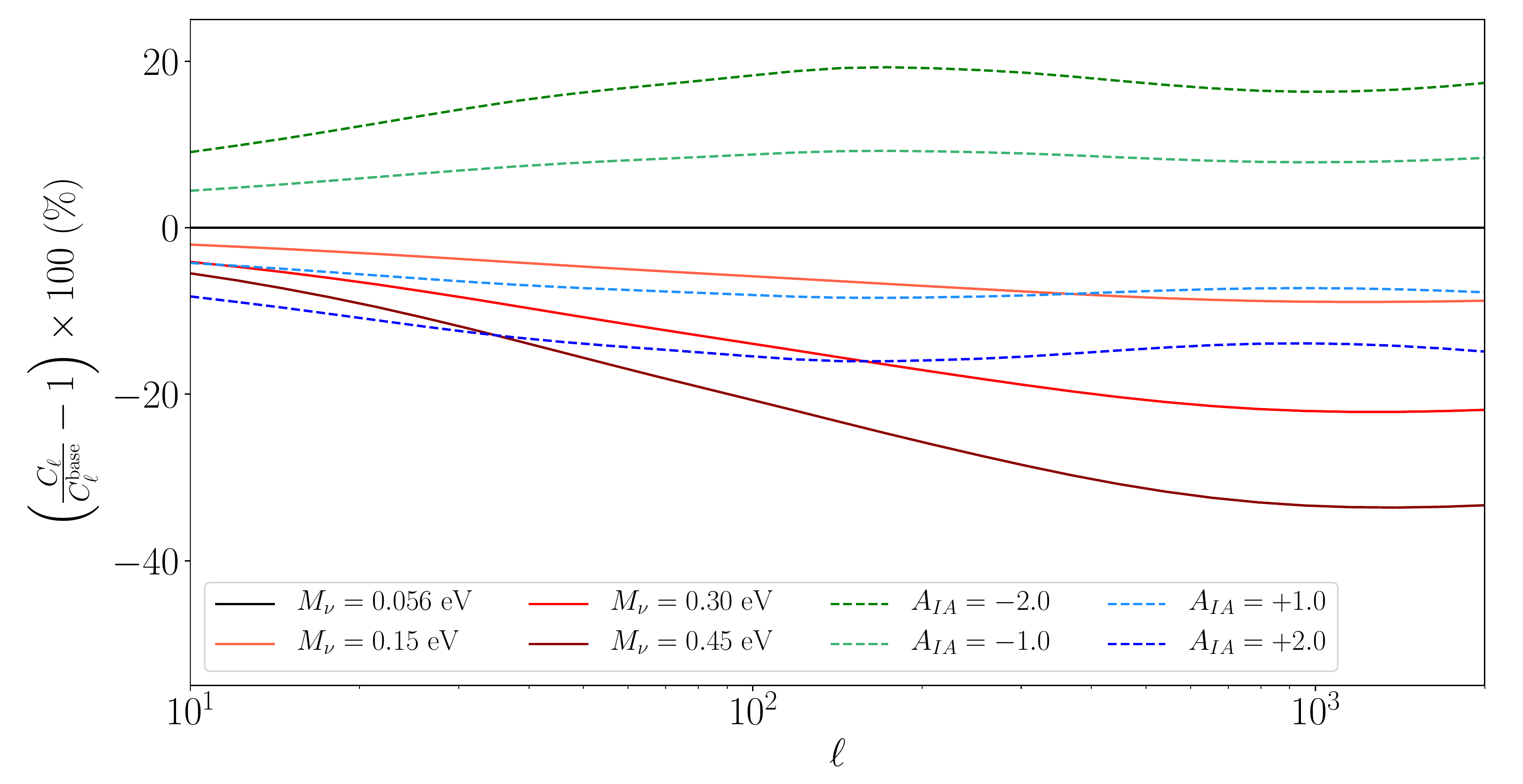}
   		  		 } 
\caption{Percentage difference on the shear power spectrum $C_\ell$ due to  an increasing neutrino mass (from red to blue) or an increasing intrinsic alignment (from green to yellow) with respect to a model with minimal neutrino mass and no intrinsic alignment. We assume here a single tomographic bin.}
\label{fig:Mnu_vs_IA_plot}
\end{figure}
\FloatBarrier

In Figure \ref{fig:Mnu_vs_IA_plot} we plot the relative difference on the shear power spectrum (we consider a single tomographic bin for simplicity) of models with different neutrino masses and models with different intrinsic alignment parameter with respect to a model with minimal neutrino mass and $A_\mathrm{IA} = 0$.
Intrinsic alignment can either enhance (if $A_\mathrm{IA}<0$) or damp (if $A_\mathrm{IA}>0$) the signal at all multipoles, and this effect may in principle mimic the neutrinos and introduce a possible degeneracy with $M_\nu$.  

We perform the MCMC with the usual method but this time setting the $z_\mathrm c$ parameter to a fixed value of 2.
We choose to do so because none of the runs of the previous analysis (Section \ref{sec:feedback_nu}) was able to constrain such parameter, due to the weak dependence of the shear spectra on it (see the bottom right panel of Figure \ref{fig:Cl_nu_feed}).
Therefore, we will have again 4 free parameters to fit: $M_\nu, \ \log M_\mathrm c, \ \eta_\mathrm b, \ A_\mathrm{IA}$.
We use the same grid of parameter values of the previous Section plus $A_\mathrm{IA} = 1.3, -1.3$.

\begin{figure}[!t]
	 \makebox[1\textwidth][c]
{\includegraphics[width=.85\textwidth]{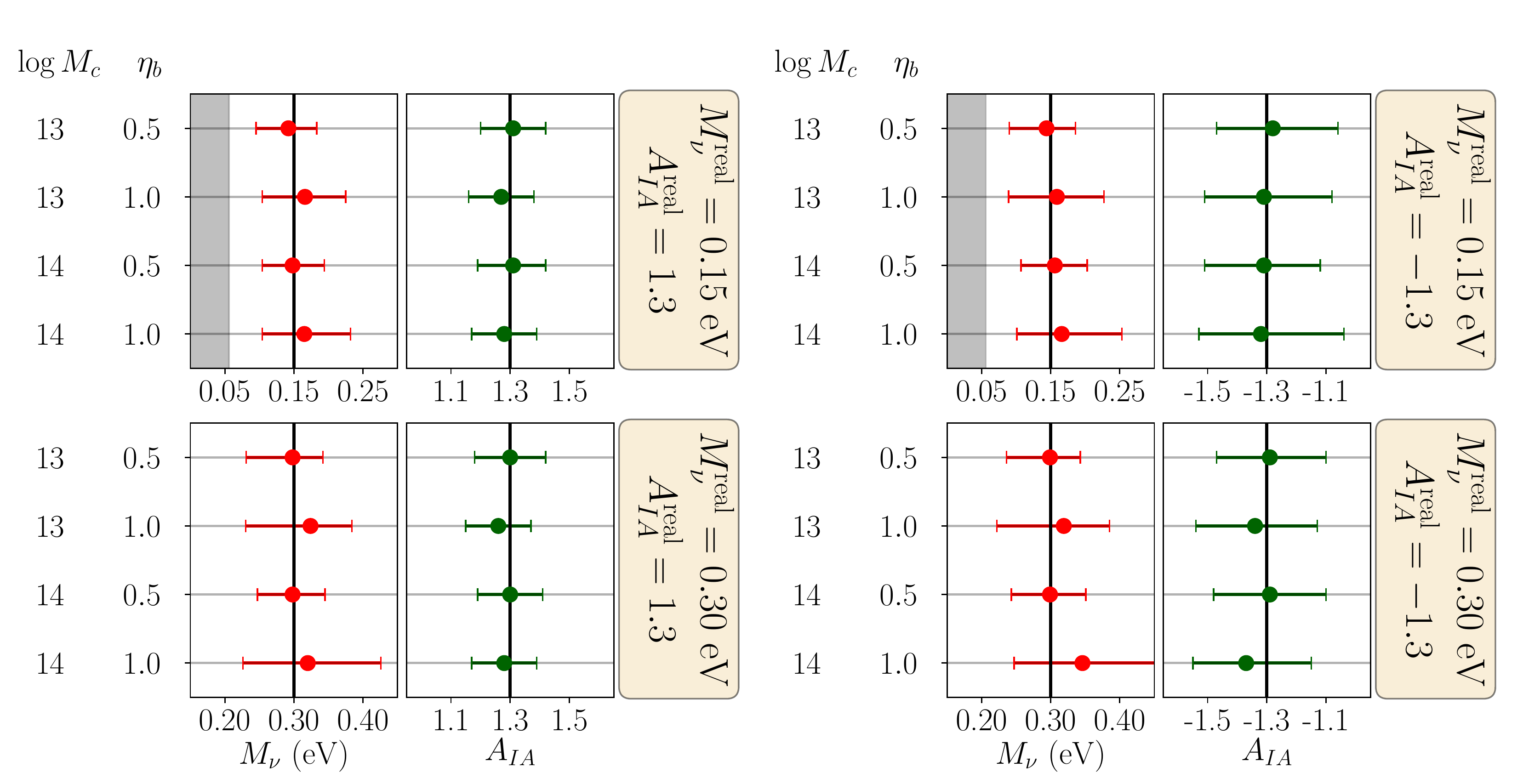}} 
\caption{Results obtained from the 16 different runs for a cosmic shear survey including the intrinsic alignment effect. For clarity we have separated the runs with same neutrino mass and intrinsic alignment parameters.
The black vertical lines represent the true input value, the error-bars mark the 68\% confidence level for neutrino mass (red) and intrinsic alignment parameter $A_\mathrm{IA}$ (green).
The left columns report the values of the feedback parameters used to generate mock data ($z_\mathrm c$ has been set to 2).
The grey shaded area is forbidden by the solar neutrino experiments.}
\label{fig:Mnu_vs_IA_results}
\end{figure}

The results for the main parameters of interest are listed in Table \ref{table:A_IA}.
For a clearer view, Figure \ref{fig:Mnu_vs_IA_results} reports the results for what concerns neutrino mass and the intrinsic alignment parameter in all the 16 different parameter sets.
We grouped the results according to the value of the input neutrino mass and the sign of $A_\mathrm{IA}$.
We see that once again we are able to recover the right input parameters.
For the neutrino mass, the maximum difference with respect to the true value is $\sim 0.46\sigma$, while for $A_\mathrm{IA}$ is $\sim 0.3\sigma$.
Again, we stress that we are assuming a perfect knowledge of the functional form and parameter values for baryonic and neutrino effects as well as the intrinsic alignment model, which in principle may be much more complicated than what we assumed.

\begin{table}[t]
\center
  \renewcommand{\arraystretch}{1.5}
  \resizebox{.8\columnwidth}{!}{
\begin{tabular}{cccc|cccc}
\multicolumn{4}{c|}{ Input parameters}  & \multicolumn{4}{c}{ $C_\ell$} \\
  \hline
$M_\nu$ & $\logMc$ & $\etab$ & $A_\mathrm{IA}$ &  $M_\nu \ [\textrm{eV}]$ & $\logMc$ & $\etab$ & $A_\mathrm{IA}$ \\
  \hline
  \midrule
  0.15 &  13 &   0.5 & 1.3 &
  $\mathbf{0.142^{+0.041}_{-0.047}}$ & $<13.4$ & $<7.5$ & $1.31^{+0.11}_{-0.11}$ \tabularnewline
  0.15 &  13 &   0.5 & -1.3 &
  $\mathbf{0.144^{+0.042}_{-0.054}}$ & $<13.6$ & $<5.0$ & $-1.28^{+0.22}_{-0.19}$ \tabularnewline
    0.15 &  13 &   1.0 & 1.3 &
  $\mathbf{0.166^{+0.059}_{-0.062}}$ & $<13.3$ & $<5.4$ & $1.27^{+0.11}_{-0.11}$ \tabularnewline
    0.15 &  13 &   1.0 & -1.3 &
    $\mathbf{0.159^{+0.068}_{-0.070}}$ & $<13.3$ & $<5.7$ & $-1.31^{+0.23}_{-0.20}$ \tabularnewline

0.15 &  14 &  0.5 & 1.3 &
  $\mathbf{0.148^{+0.046}_{-0.044}}$ & $14.0^{+0.3}_{-0.3}$ & $0.5^{+0.1}_{-0.1}$ &  $1.31^{+0.11}_{-0.12}$ \tabularnewline
0.15 &  14 &   0.5 & -1.3 &
  $\mathbf{0.156^{+0.047}_{-0.049}}$ & $14.0^{+0.3}_{-0.3}$ & $0.5^{+0.1}_{-0.1}$ &  $-1.31^{+0.19}_{-0.20}$ \tabularnewline
0.15 &  14 &   1.0 & 1.3 &
    $\mathbf{0.165^{+0.067}_{-0.061}}$ & $13.9^{+0.4}_{-0.4}$ & $1.0^{+0.2}_{-0.2}$ &  $1.28^{+0.11}_{-0.11}$ \tabularnewline
0.15 &  14 &   1.0 & -1.3 &
  $\mathbf{0.166^{+0.087}_{-0.065}}$ & $13.9^{+0.5}_{-0.5}$ & $1.0^{+0.2}_{-0.4}$ &  $-1.32^{+0.28}_{-0.21}$ \tabularnewline

  \hline
0.30 &  13 &   0.5 & 1.3 &
  $\mathbf{0.298^{+0.044}_{-0.067}}$ & $<13.8$ & $<3.1$ & $1.30^{+0.12}_{-0.12}$ \tabularnewline
0.30 &  13 &   0.5 & -1.3 &
  $\mathbf{0.299^{+0.044}_{-0.063}}$ & $<13.5$ & $<9.6$ & $-1.29^{+0.19}_{-0.18}$ \tabularnewline
0.30 &  13 &   1.0 & 1.3 &
  $\mathbf{0.324^{+0.060}_{-0.094}}$ & $<13.5$ & $<10.4$ & $1.26^{+0.11}_{-0.11}$ \tabularnewline
0.30 &  13 &   1.0 & -1.3 &
  $\mathbf{0.319^{+0.066}_{-0.097}}$ & $<13.5$ & $<3.7$ & $-1.34^{+0.21}_{-0.20}$ \tabularnewline

0.30 &  14 &   0.5 & 1.3 &
  $\mathbf{0.298^{+0.047}_{-0.051}}$ & $14.0^{+0.3}_{-0.3}$ & $0.5^{+0.1}_{-0.1}$ &  $1.30^{+0.11}_{-0.11}$ \tabularnewline
0.30 &  14 &   0.5 & -1.3 &
   $\mathbf{0.299^{+0.052}_{-0.056}}$ & $14.1^{+0.3}_{-0.3}$ & $0.5^{+0.1}_{-0.1}$ &  $-1.29^{+0.19}_{-0.19}$ \tabularnewline
 0.30 &  14 &   1.0 & 1.3 &
  $\mathbf{0.320^{+0.106}_{-0.094}}$ & $13.9^{+0.6}_{-0.8}$ & $1.0^{+0.3}_{-0.2}$ &  $1.28^{+0.11}_{-0.11}$ \tabularnewline
 0.30 &  14 &   1.0 & -1.3 &
  $\mathbf{0.346^{+0.160}_{-0.099}}$ & $14.0^{+2.7}_{-0.6}$ & $1.0^{+0.2}_{-0.9}$ &  $-1.37^{+0.22}_{-0.18}$ \tabularnewline
  \bottomrule
  \end{tabular}
}
\caption{This table shows the 68\% confidence level intervals for the parameters obtained from the analysis of cosmic shear power spectra to which the intrinsic alignment contribution has been added. See Section \ref{sec:intrinsic_alignment} for details.}
\label{table:A_IA}
\end{table}

\begin{figure}[!t]
        \centering
        \begin{subfigure}[b]{0.6\textwidth}
            \centering
            \includegraphics[width=\textwidth]{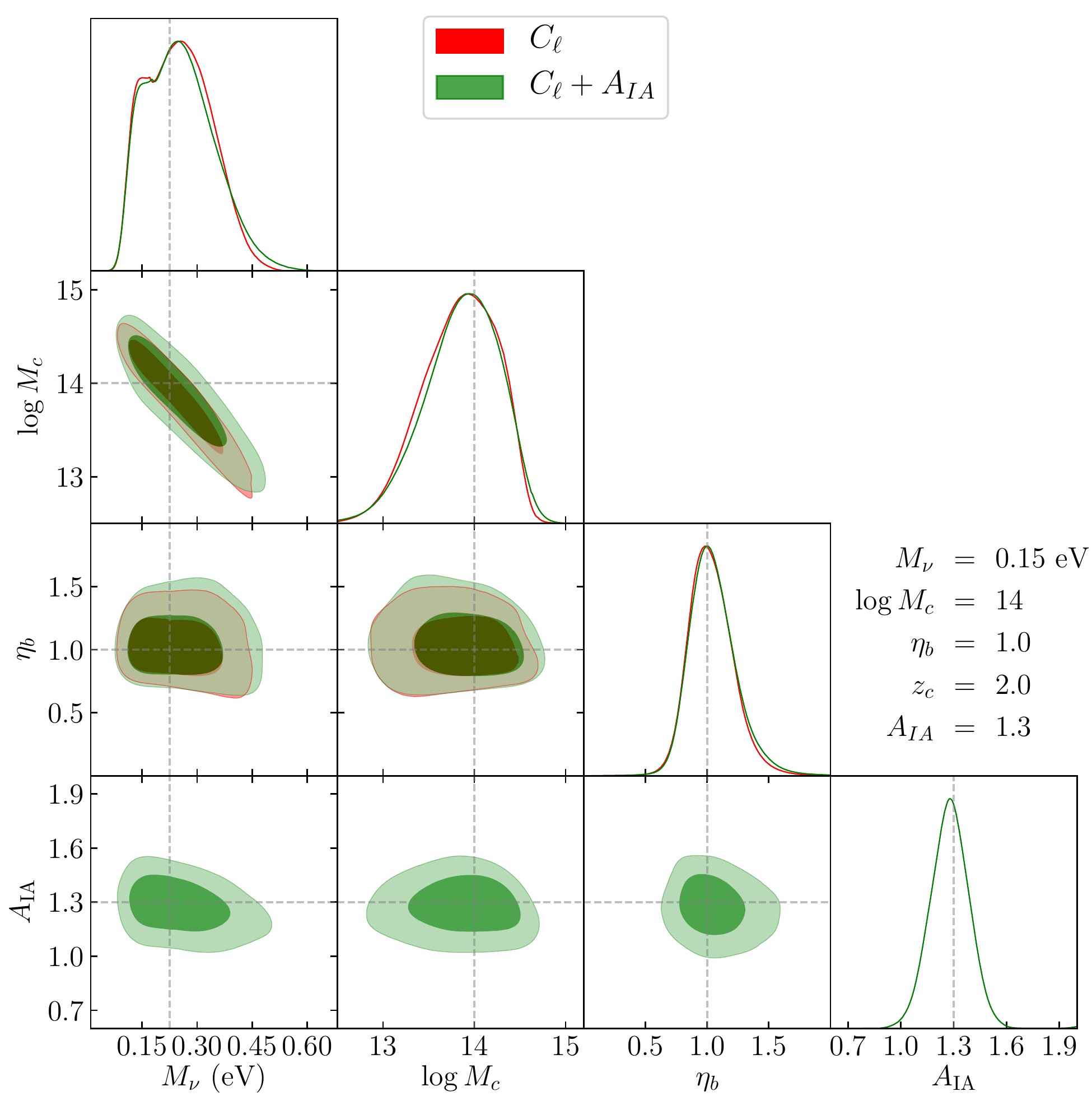}
        \end{subfigure}
        \hspace{-.7\baselineskip}
        \begin{subfigure}[b]{0.6\textwidth}  
            \centering 
            \includegraphics[width=\textwidth]{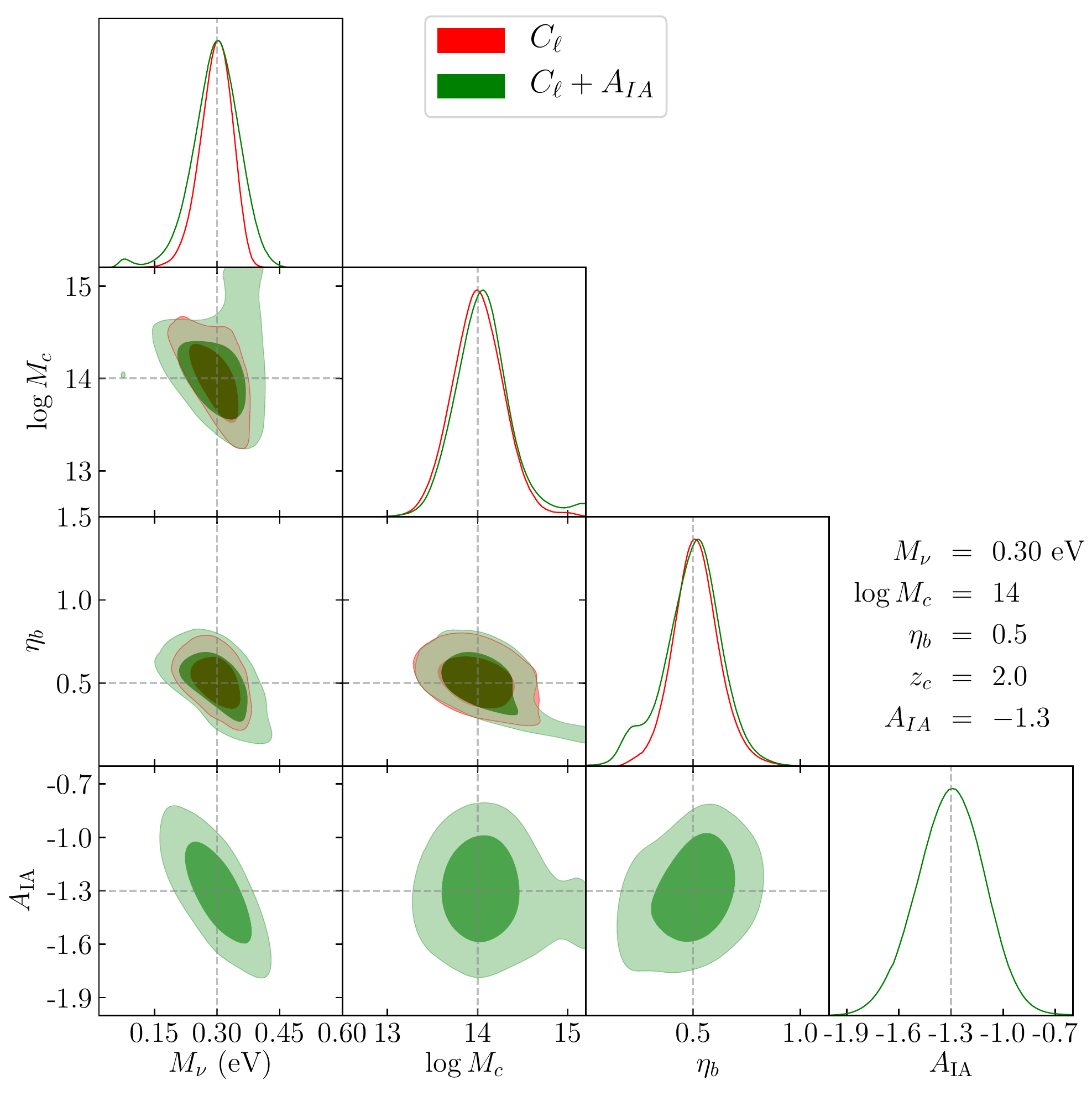}
        \end{subfigure}
\caption{Triangle plot showing 1-D and 2-D posterior PDFs for $M_\nu$, $\log M_\mathrm c$, $\eta_\mathrm b$ and $A_\mathrm{IA}$ for two of the 16 runs described in Sections \ref{sec:feedback_nu} and \ref{sec:intrinsic_alignment} whose parameters are displayed in the plot.
The red and green contours represent the results from the cosmic shear survey with and without intrinsic alignment, respectively.
The grey dashed lines show the ``true'' values, used to generate the mock data.}
\label{fig:triangle}
\end{figure}

Figure \ref{fig:triangle} shows the contour plots for two combinations of parameters (the same of Figure \ref{fig:triangle_4p}).
In green we show the $1\sigma$ and $2\sigma$ contours for the run with intrinsic alignment, while the red ones are the same contours of Figure \ref{fig:triangle_4p} (and therefore the posterior on $A_\mathrm{IA}$ is absent). The results and in particular the degeneracy patterns are rather similar: 
the only particularly pronounced degeneracy lies in the $M_\nu-\log M_\mathrm c$ plane, but it is essentially unaffected by intrinsic alignment, while those between the other feedback parameters are pretty weak.

\begin{figure}[!t]
        \centering
        \begin{subfigure}[b]{0.25\textwidth}
            \centering
            \includegraphics[width=\textwidth]{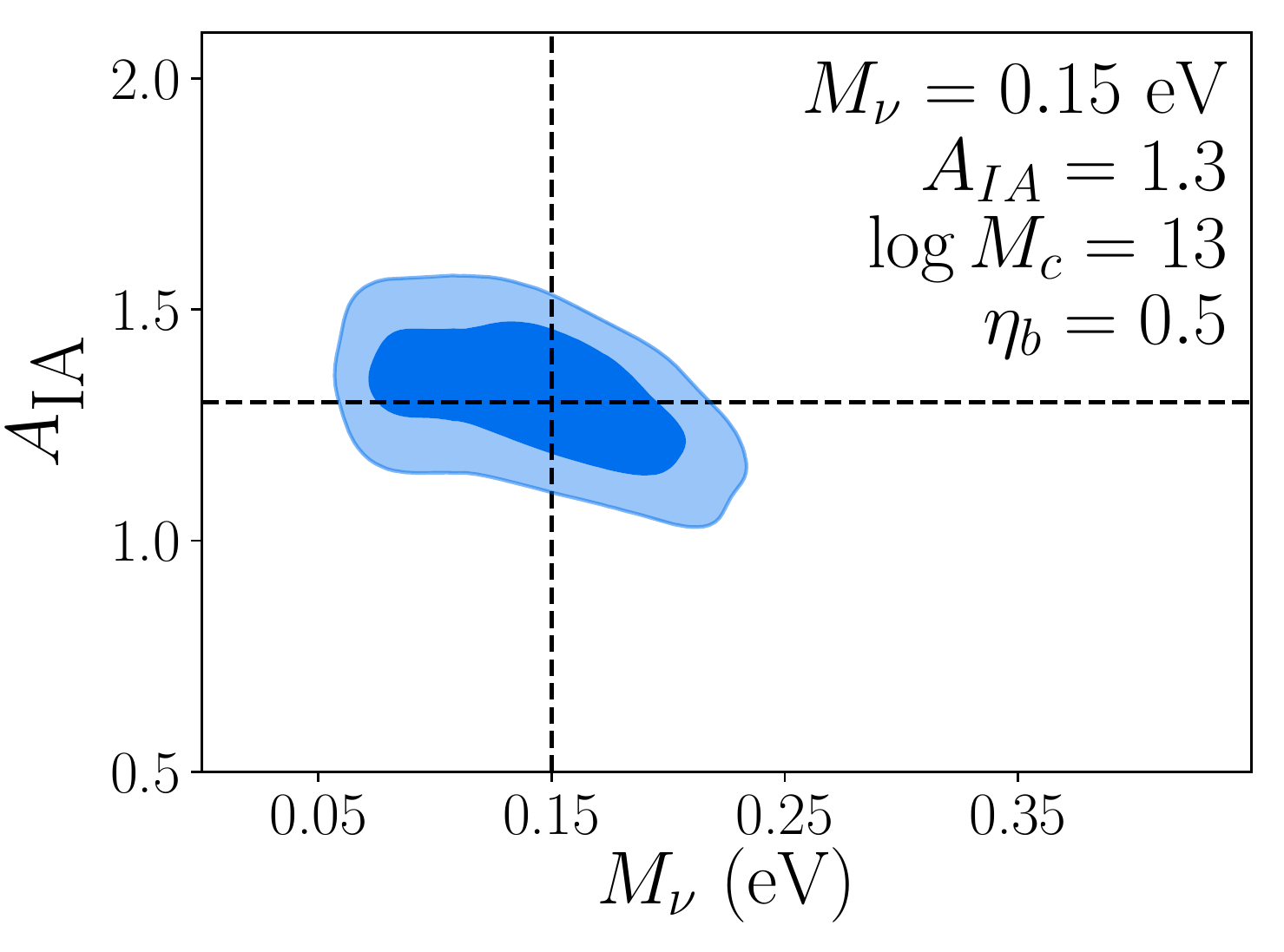}
        \end{subfigure}
        \hspace{-.7\baselineskip}
        \begin{subfigure}[b]{0.25\textwidth}  
            \centering 
            \includegraphics[width=\textwidth]{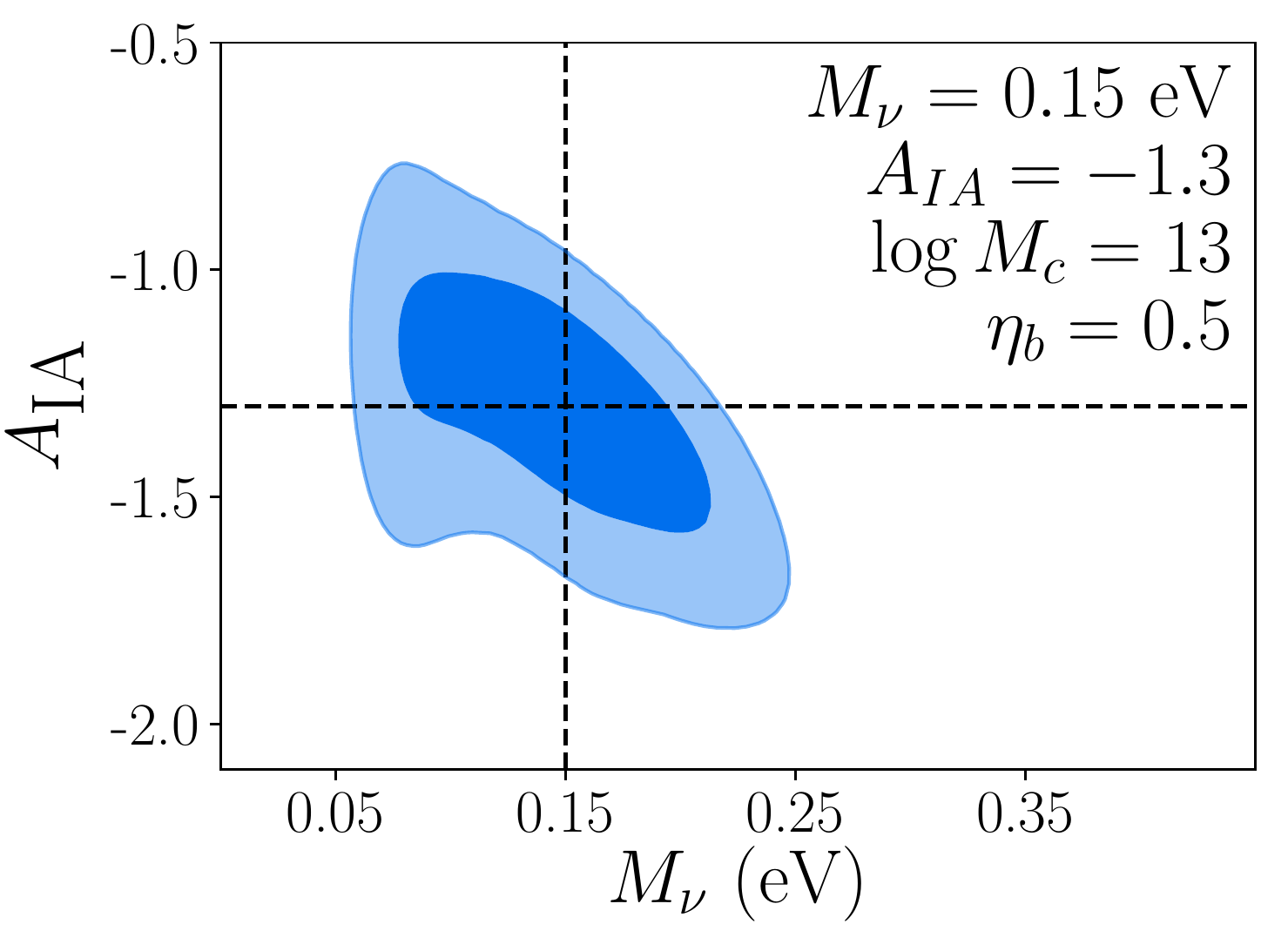}
        \end{subfigure}
        \hspace{-.7\baselineskip}
        \begin{subfigure}[b]{0.25\textwidth}
            \centering
            \includegraphics[width=\textwidth]{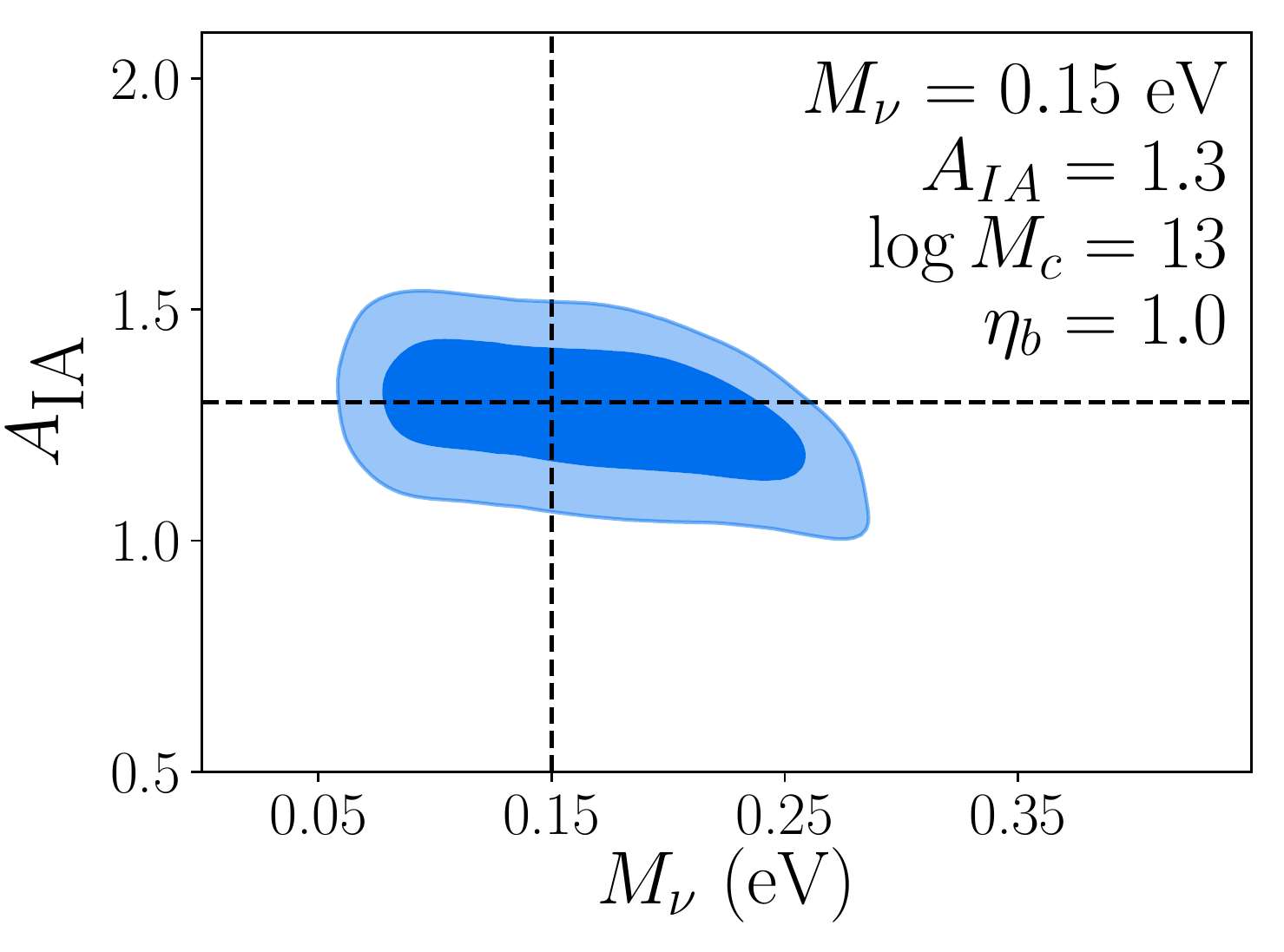}
        \end{subfigure}
        \hspace{-.7\baselineskip}
        \begin{subfigure}[b]{0.25\textwidth}  
            \centering 
            \includegraphics[width=\textwidth]{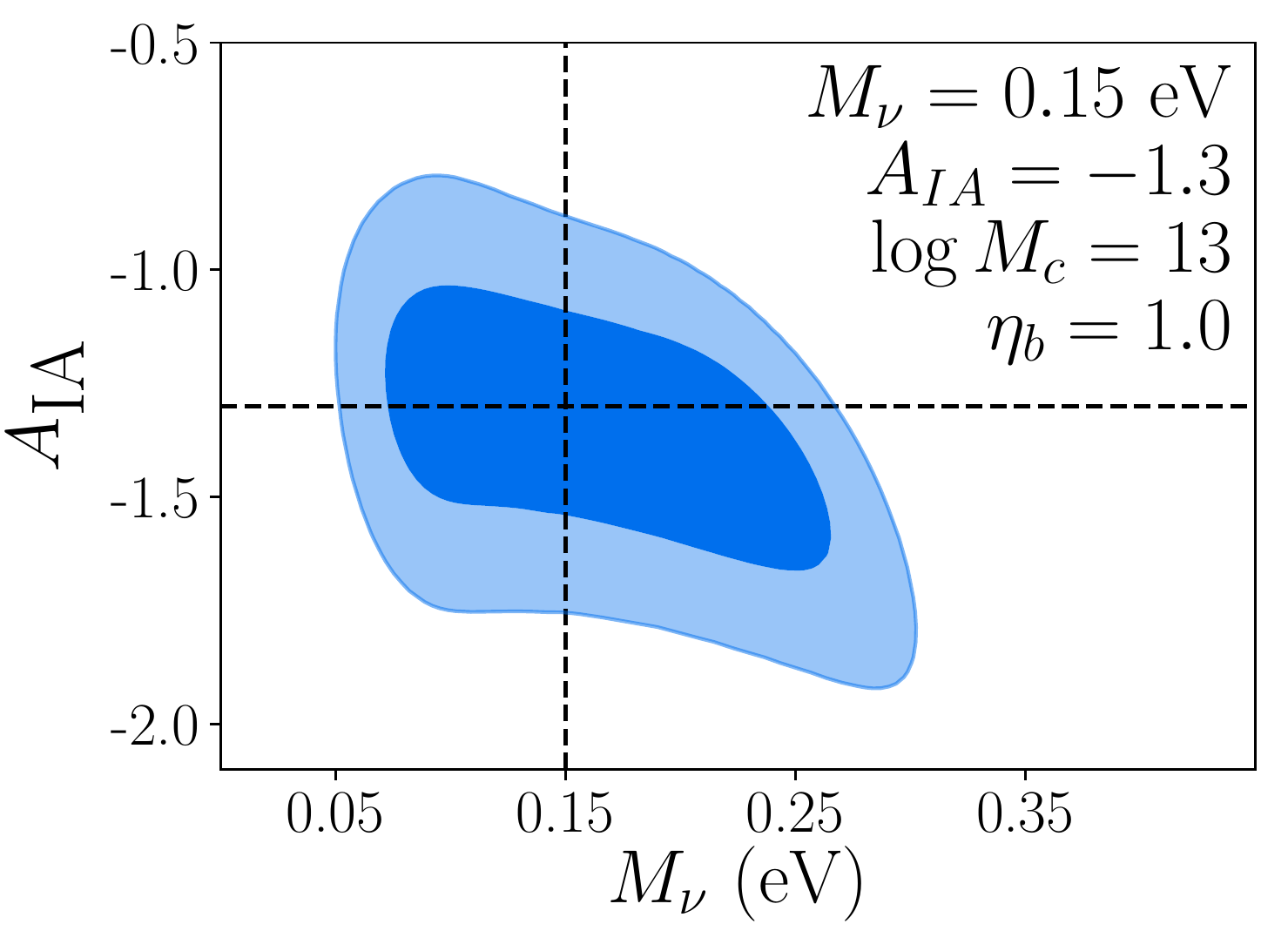}
        \end{subfigure}

        \begin{subfigure}[b]{0.25\textwidth}
            \centering
            \includegraphics[width=\textwidth]{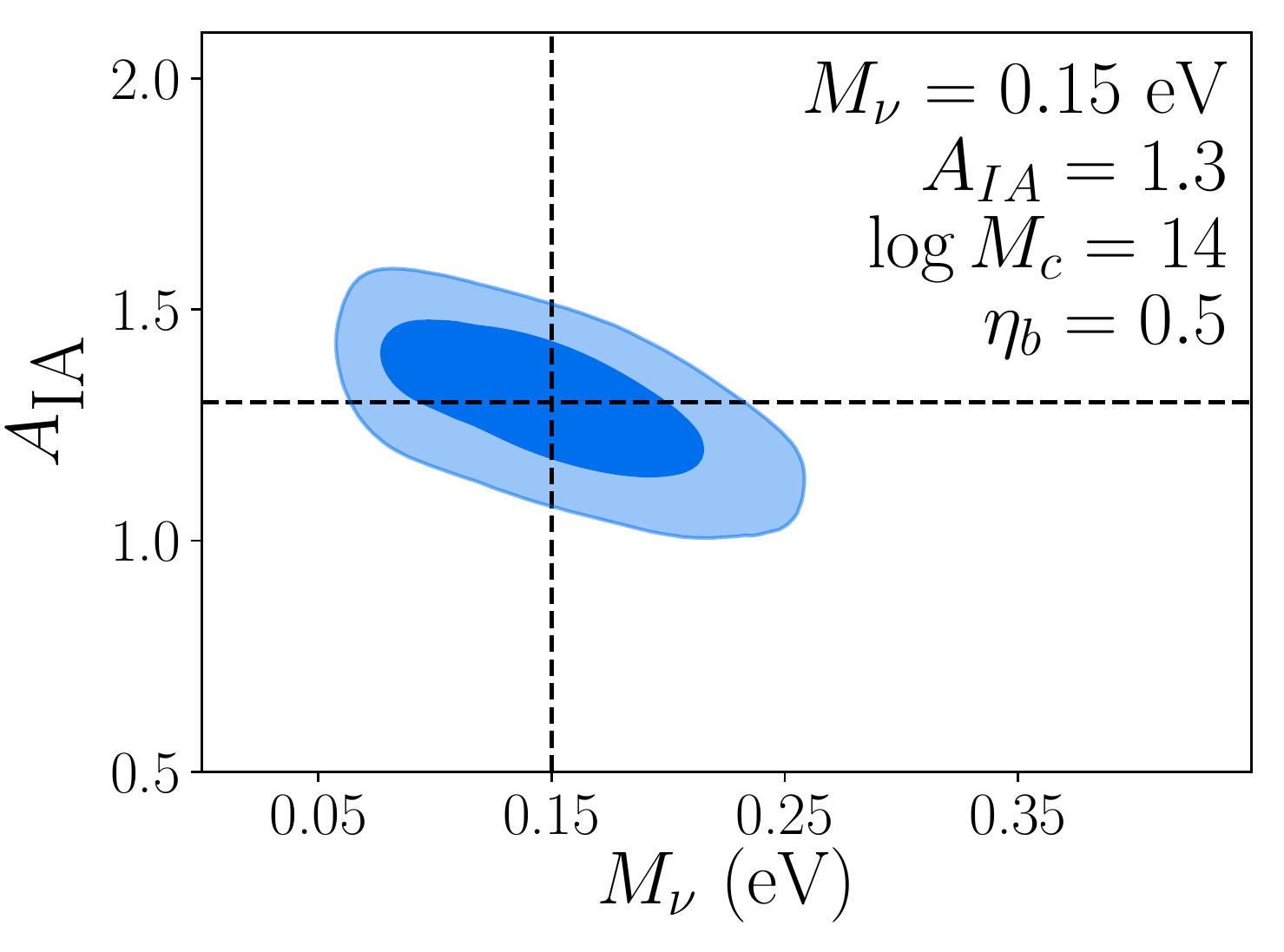}
        \end{subfigure}
        \hspace{-.7\baselineskip}
        \begin{subfigure}[b]{0.25\textwidth}  
            \centering 
            \includegraphics[width=\textwidth]{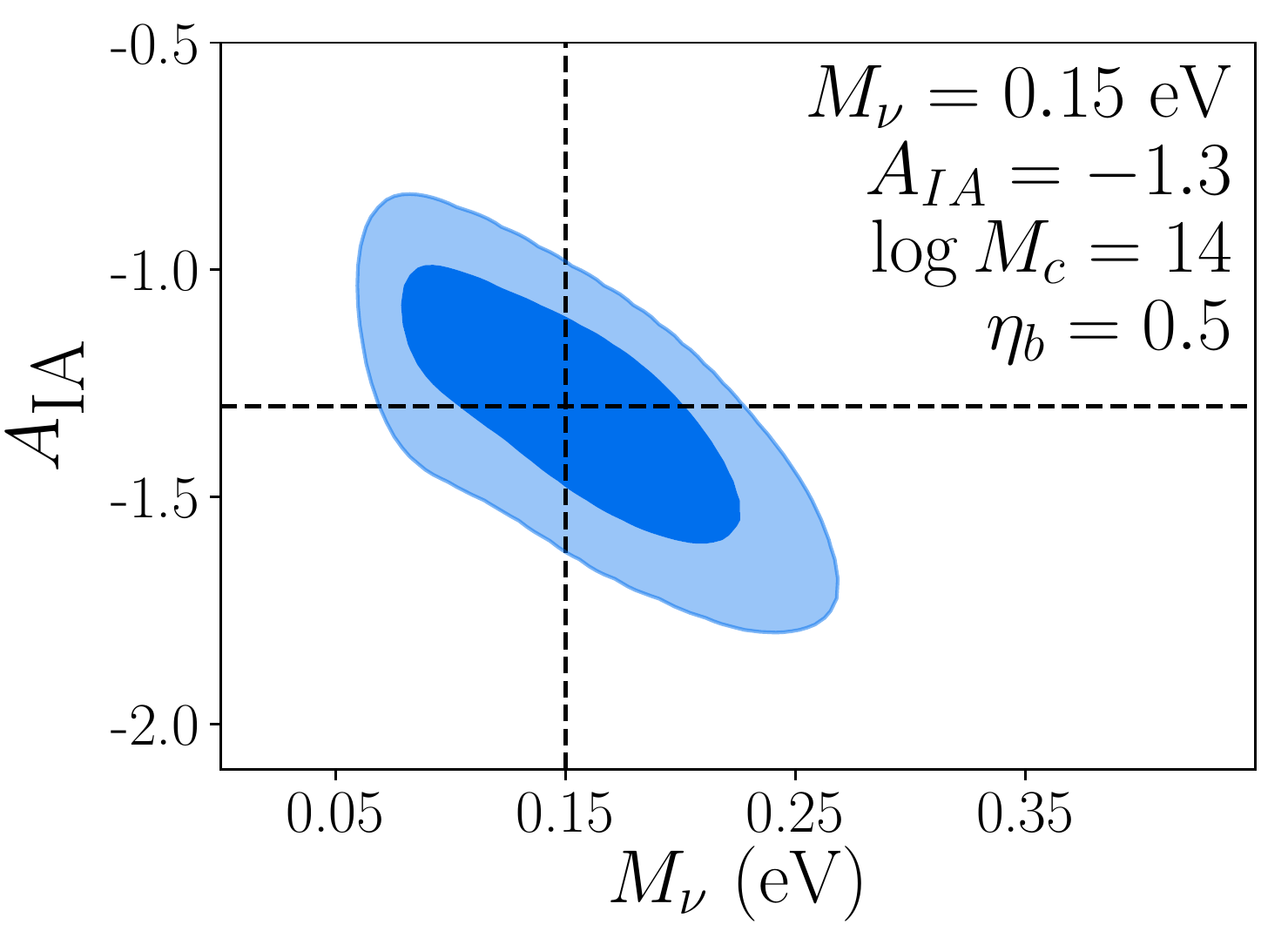}
        \end{subfigure}
        \hspace{-.7\baselineskip}
        \begin{subfigure}[b]{0.25\textwidth}
            \centering
            \includegraphics[width=\textwidth]{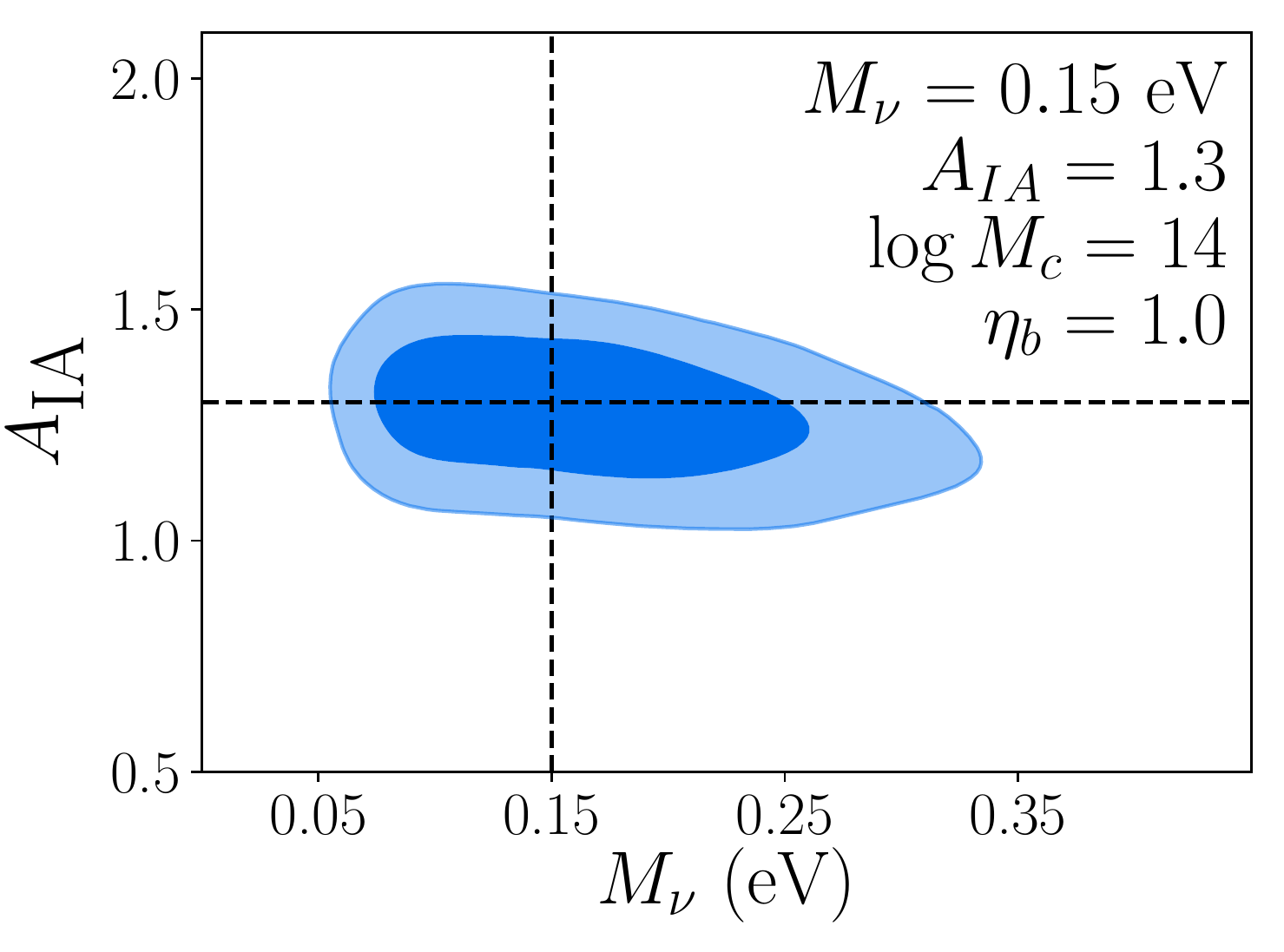}
        \end{subfigure}
        \hspace{-.7\baselineskip}
        \begin{subfigure}[b]{0.25\textwidth}  
            \centering 
            \includegraphics[width=\textwidth]{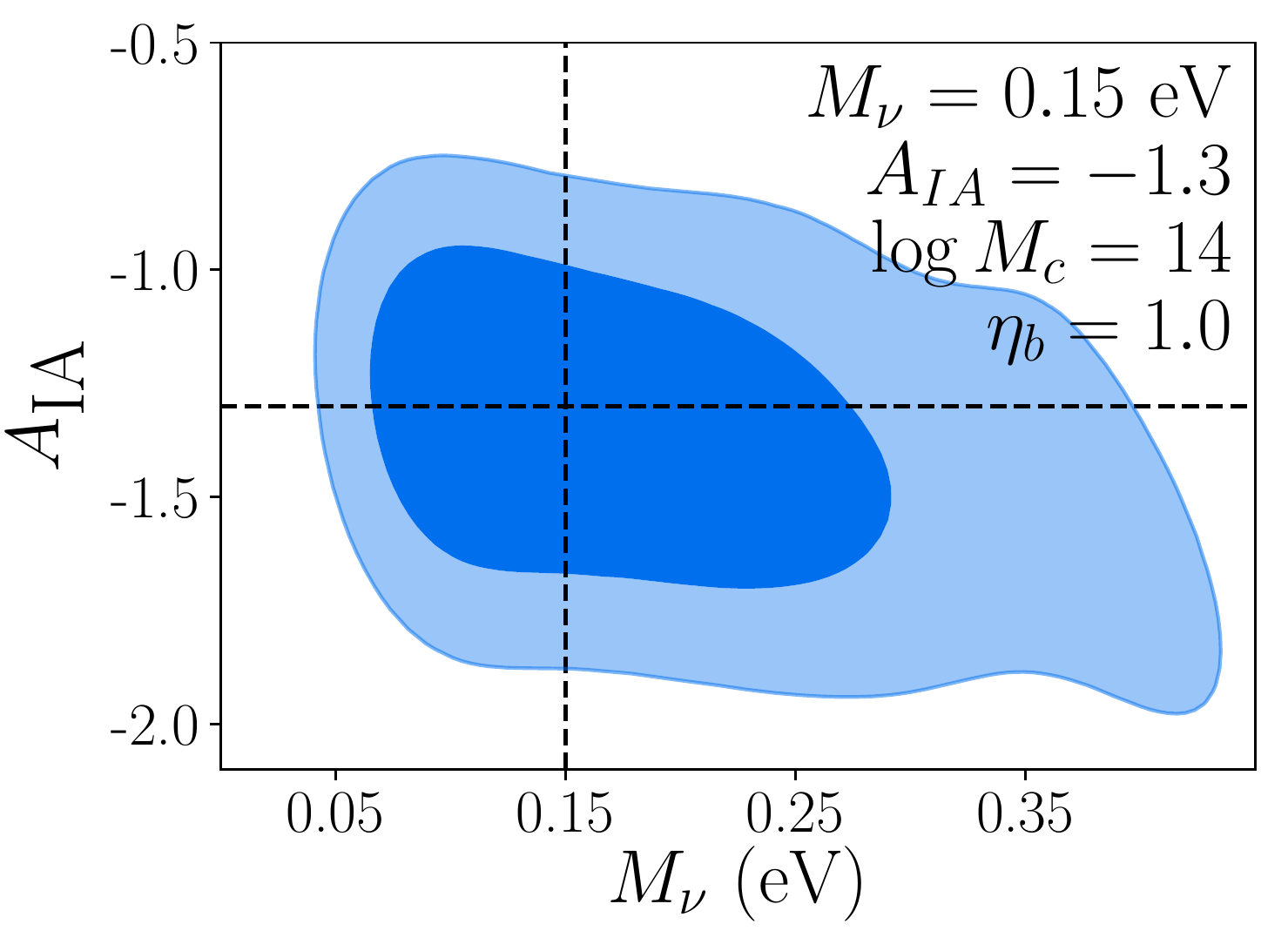}
        \end{subfigure}

        \begin{subfigure}[b]{0.25\textwidth}
            \centering
            \includegraphics[width=\textwidth]{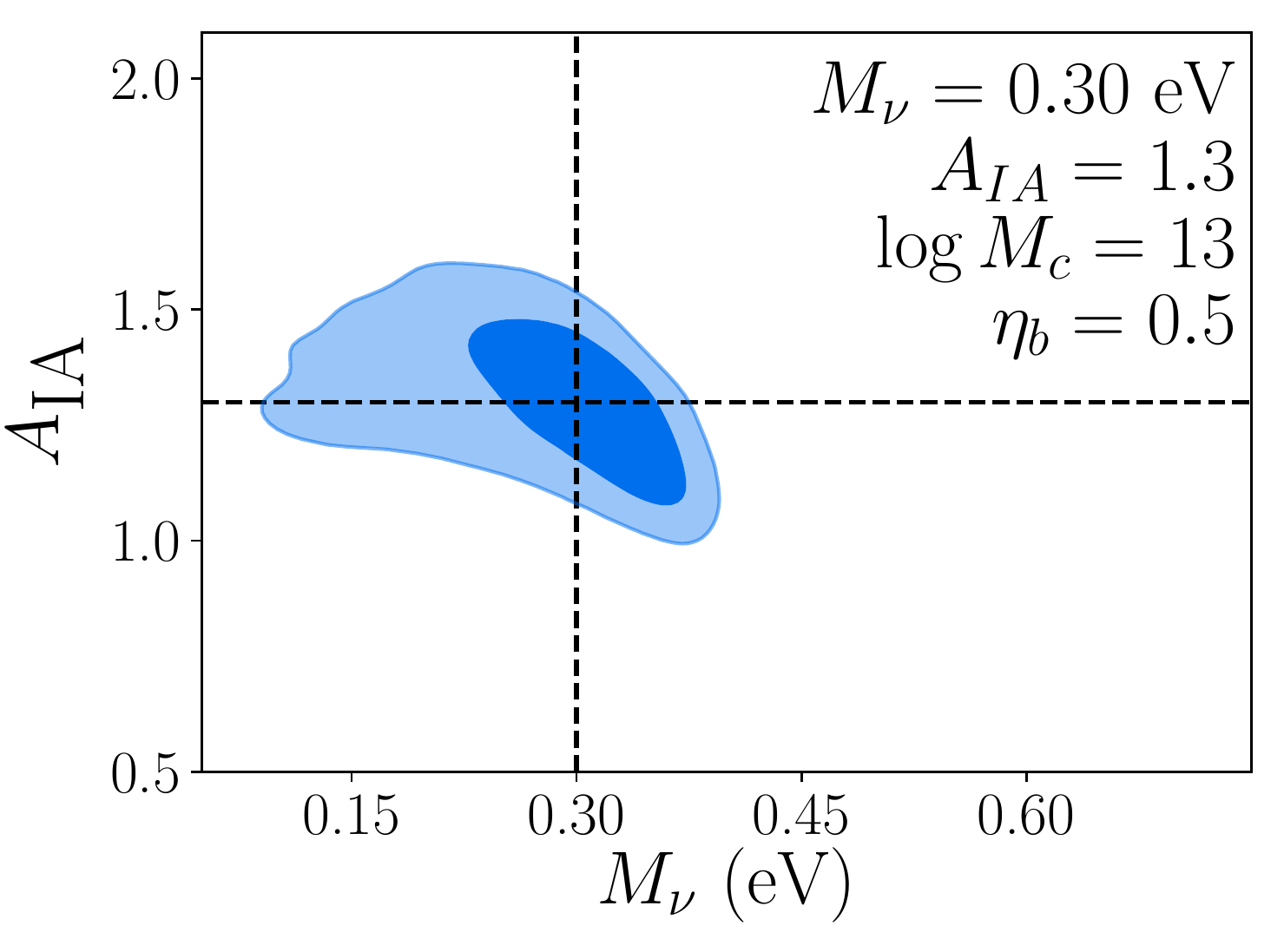}
        \end{subfigure}
        \hspace{-.7\baselineskip}
        \begin{subfigure}[b]{0.25\textwidth}  
            \centering 
            \includegraphics[width=\textwidth]{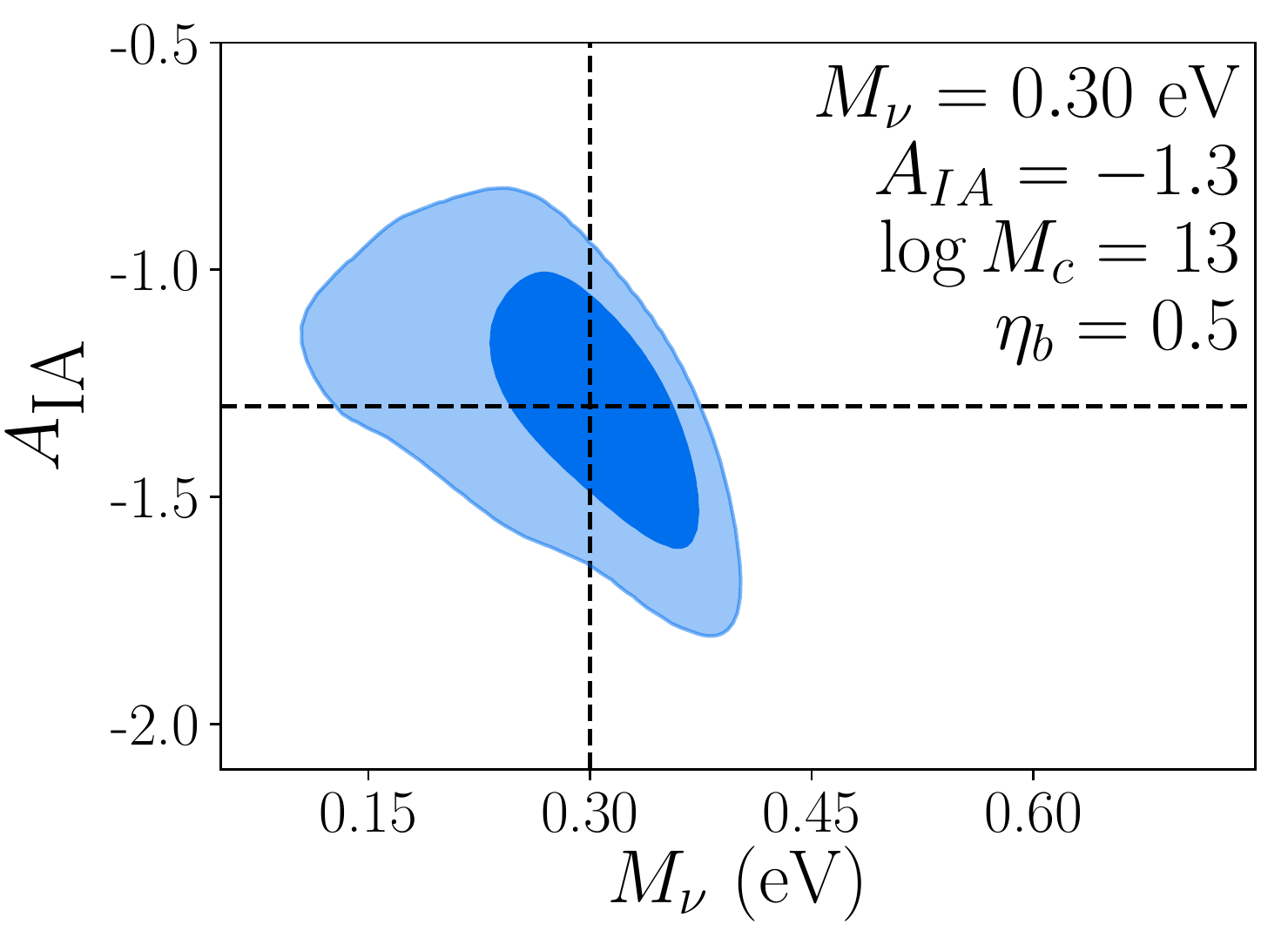}
        \end{subfigure}
        \hspace{-.7\baselineskip}
        \begin{subfigure}[b]{0.25\textwidth}
            \centering
            \includegraphics[width=\textwidth]{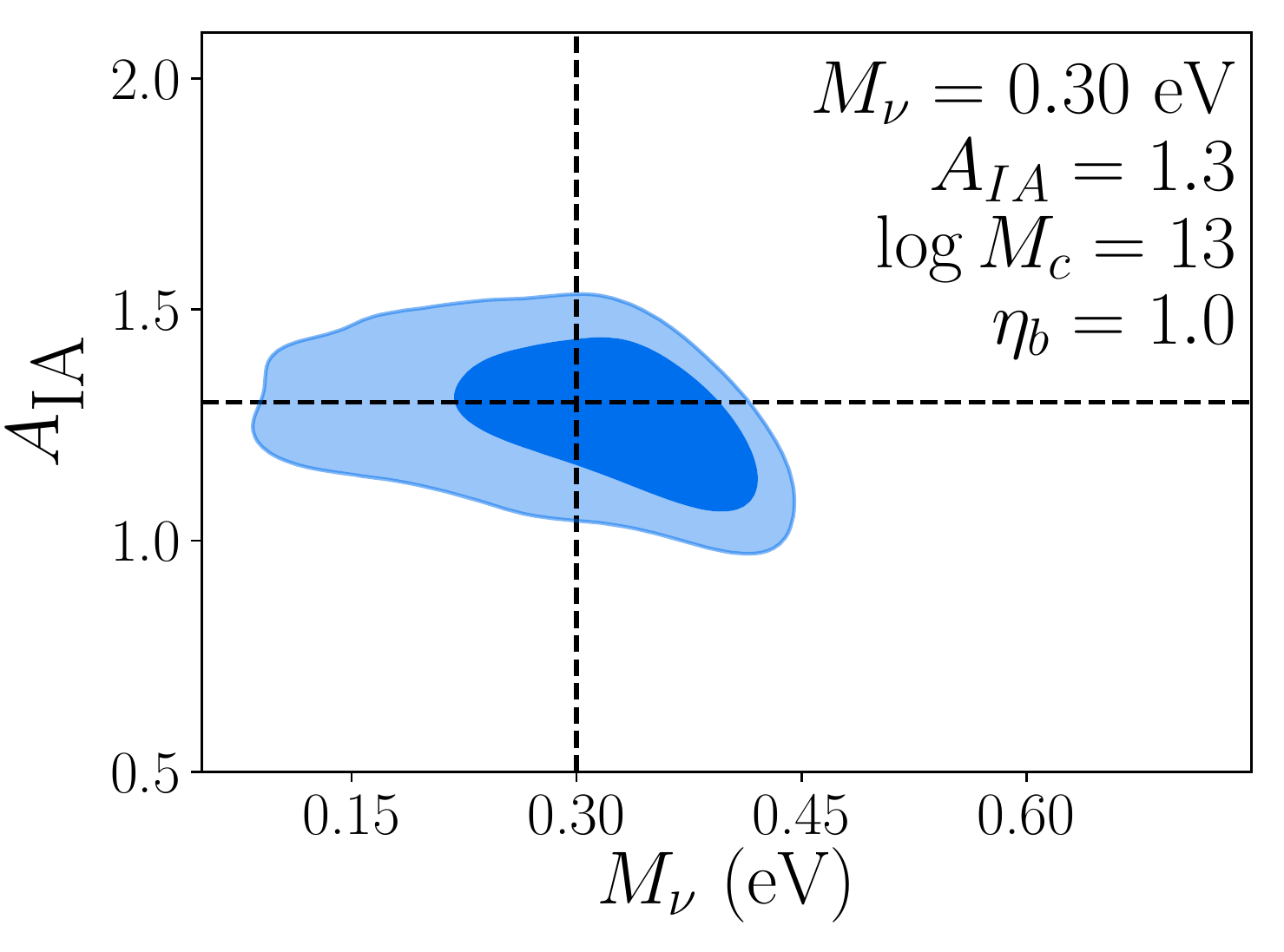}
        \end{subfigure}
        \hspace{-.7\baselineskip}
        \begin{subfigure}[b]{0.25\textwidth}  
            \centering 
            \includegraphics[width=\textwidth]{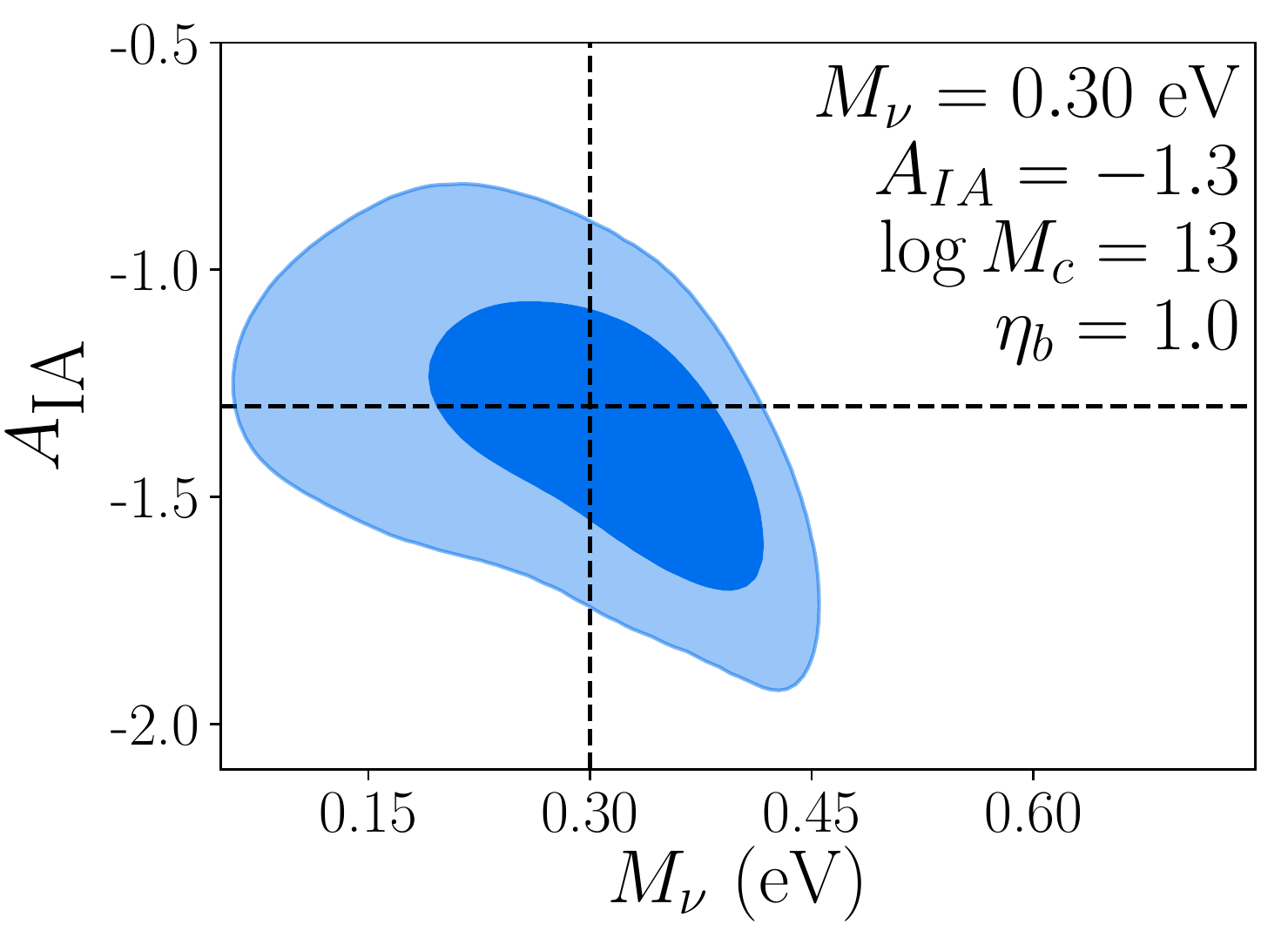}
        \end{subfigure}

        \begin{subfigure}[b]{0.25\textwidth}
            \centering
            \includegraphics[width=\textwidth]{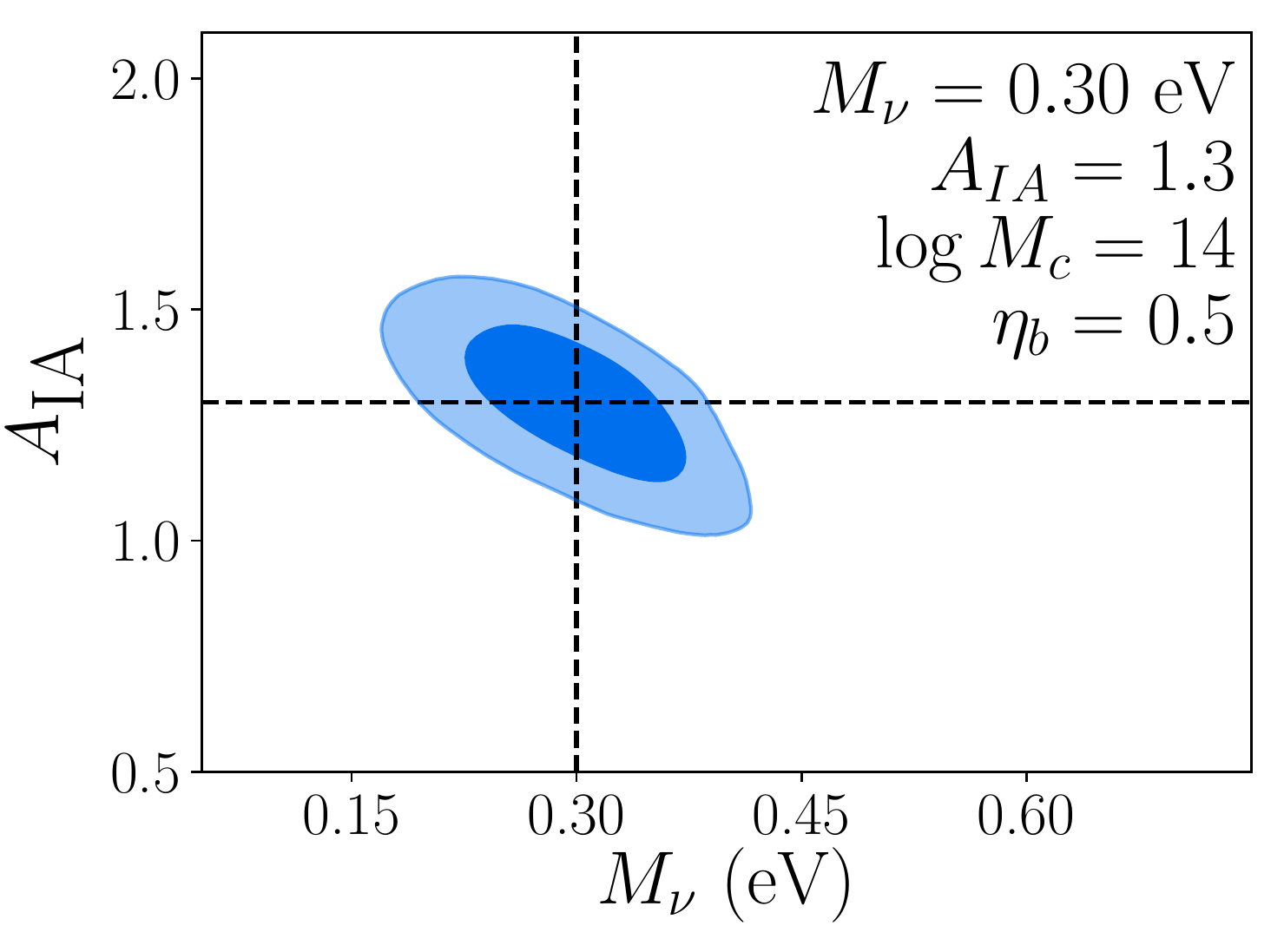}
        \end{subfigure}
        \hspace{-.7\baselineskip}
        \begin{subfigure}[b]{0.25\textwidth}  
            \centering 
            \includegraphics[width=\textwidth]{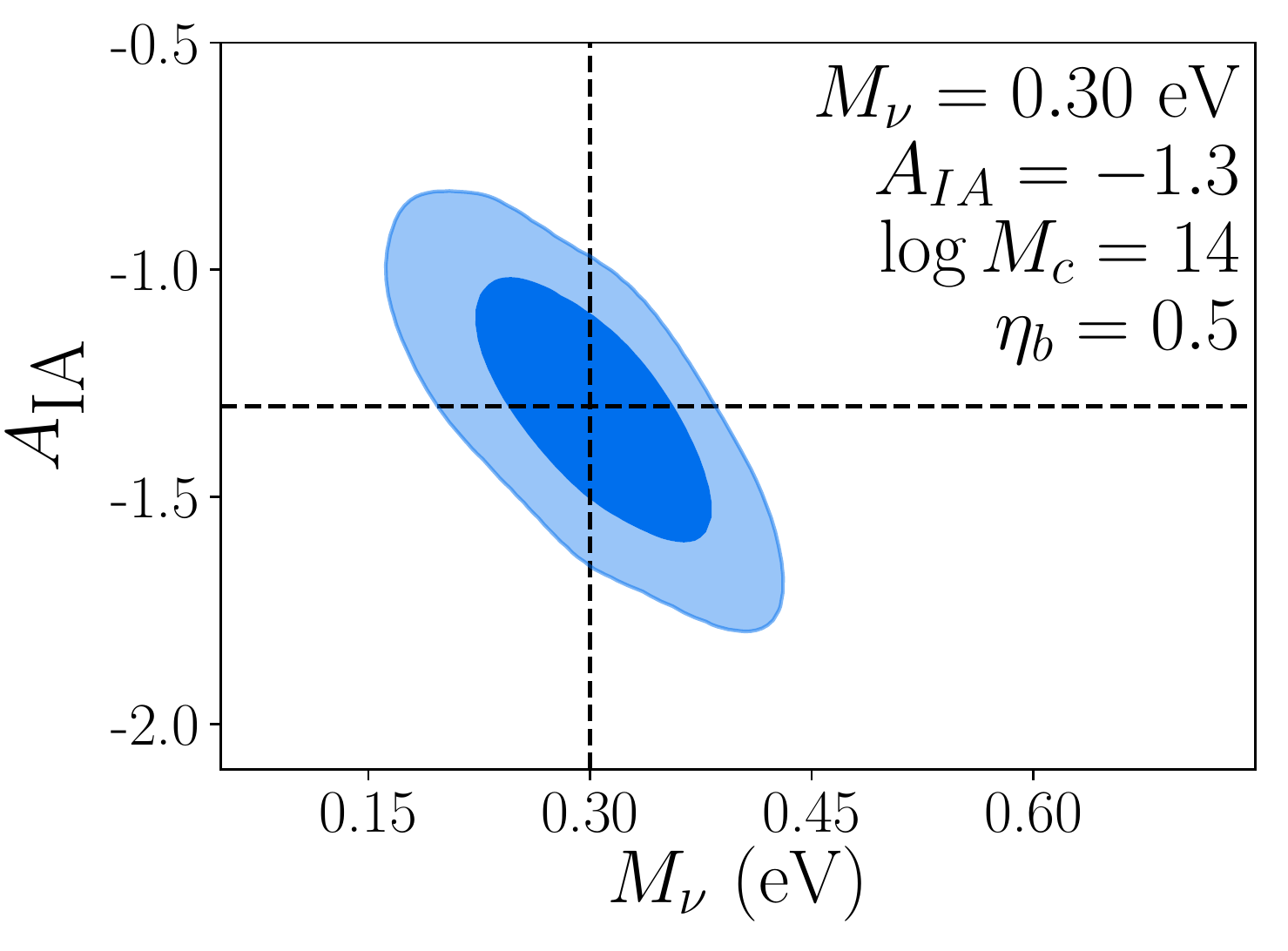}
        \end{subfigure}
        \hspace{-.7\baselineskip}
        \begin{subfigure}[b]{0.25\textwidth}
            \centering
            \includegraphics[width=\textwidth]{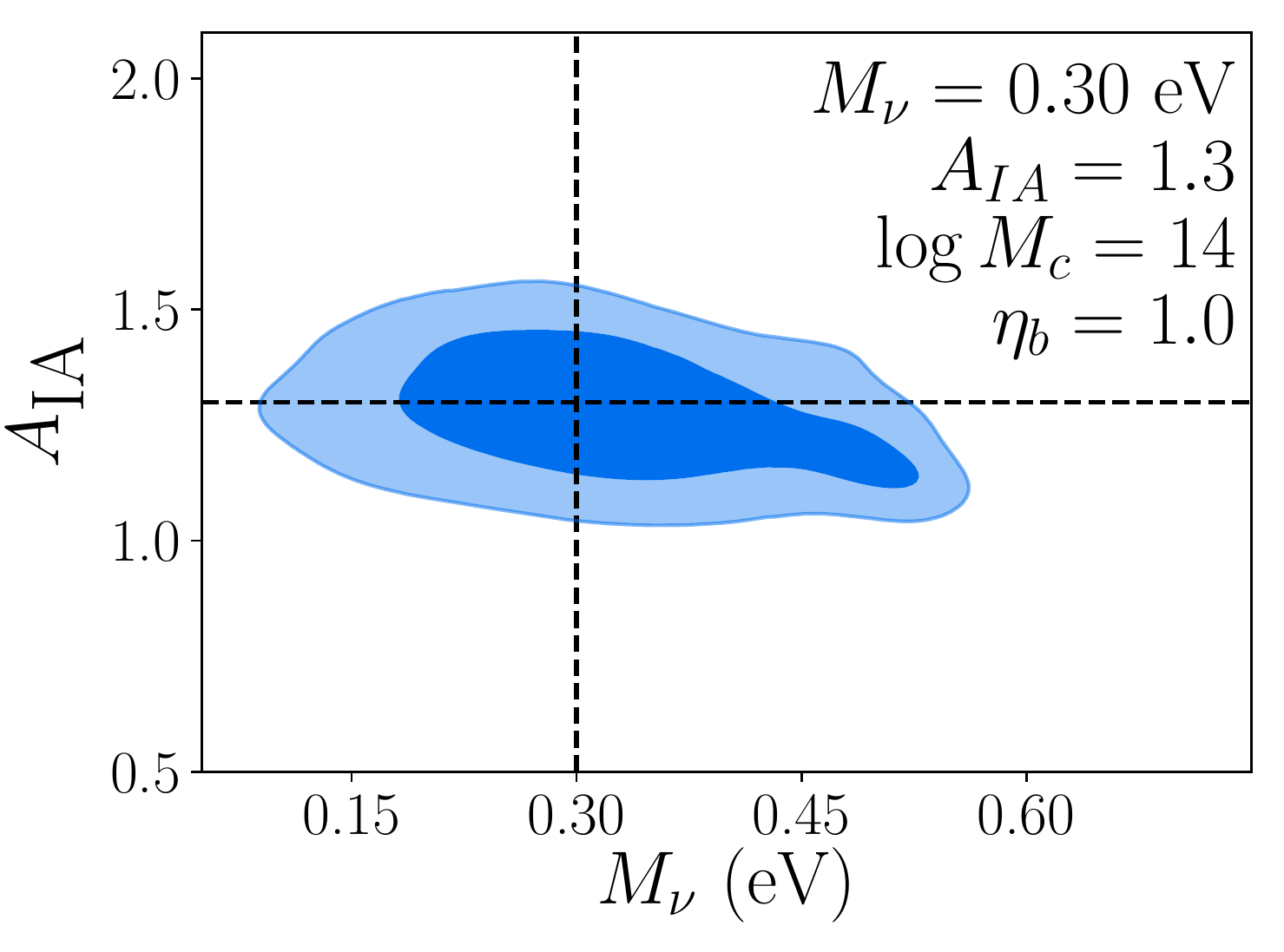}
        \end{subfigure}
        \hspace{-.7\baselineskip}
        \begin{subfigure}[b]{0.25\textwidth}  
            \centering 
            \includegraphics[width=\textwidth]{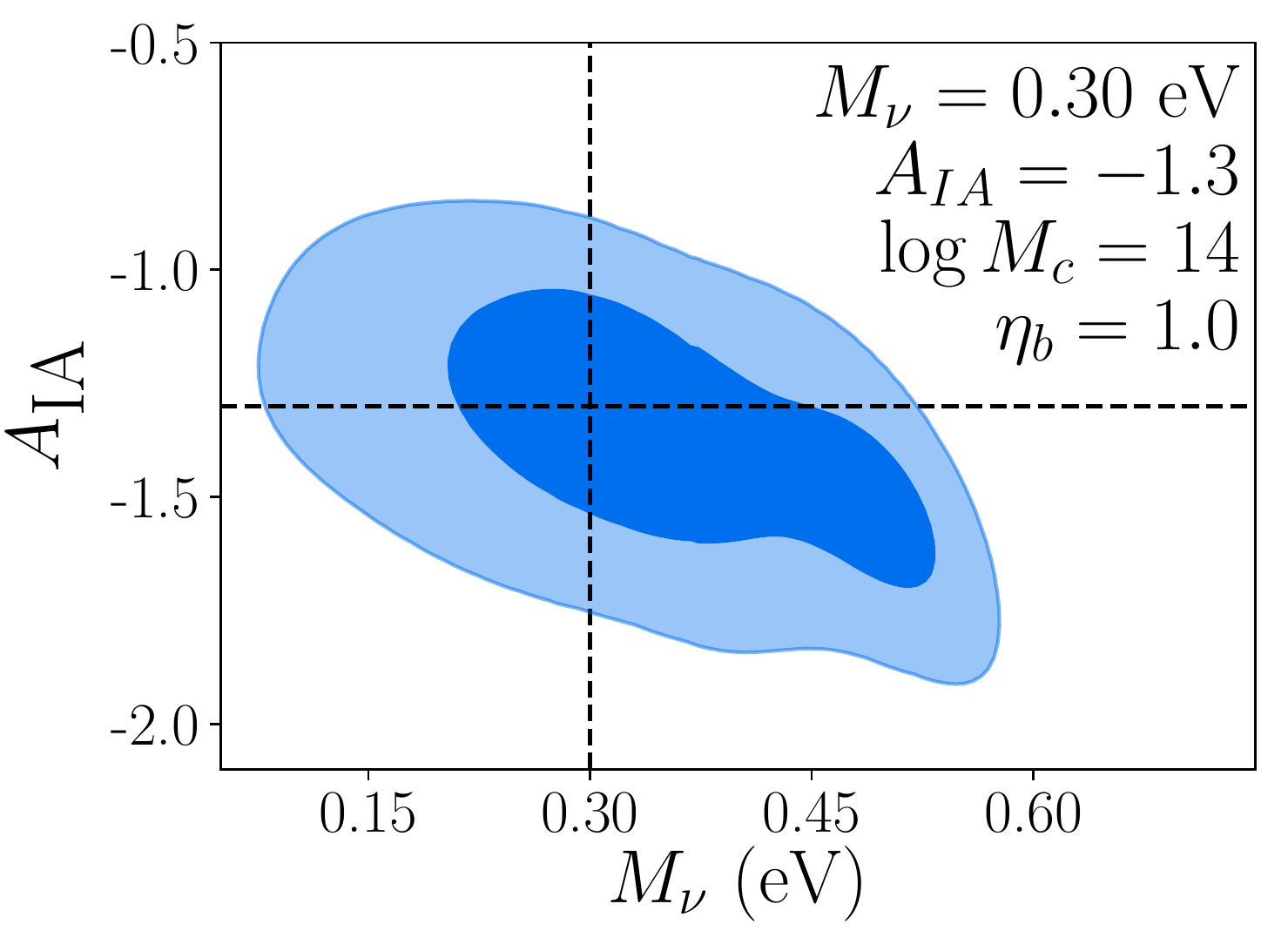}
        \end{subfigure}
        \vspace{\baselineskip}
        \caption{This picture shows the degeneracy between neutrino mass and the intrinsic alignment parameter for the 16 different cases analysed in Section \ref{sec:intrinsic_alignment}.
The top plots have $M_\nu = 0.15$ eV, while the bottom ones have $M_\nu = 0.3$ eV; odd columns have $A_\mathrm{IA} = 1.3$, even columns have $A_\mathrm{IA} = -1.3$ (the dashed lines help the view in marking the true value). The parameters of the set are written inside each panel.
The contour lines shown are 68\% and 95\% confidence level, while the dashed black lines show the true values for the parameters. It is clearly visible that in some cases the degeneracy between the two parameters is totally absent, but even where is present it will be likely not to bias the measurement on neutrino mass.}
        \label{fig:Mnu_vs_AIA}
\end{figure}

An interesting point to discuss concerns the degeneracy between $M_\nu$ and $A_\mathrm{IA}$.
Figure \ref{fig:Mnu_vs_AIA} shows the 1-$\sigma$ and 2-$\sigma$ contour lines in the $M_\nu-A_\mathrm{IA}$ plane for the 16 different parameter sets we used.
We see that there seems to be a degeneracy pattern which is weak for positive $A_\mathrm{IA}$ (odd columns) and stronger for negative $A_\mathrm{IA}$ (even columns).
In particular we find an anti-correlation between the two parameters.
This is expected, since $M_\nu$ lowers the signal at almost all multipoles (see Figure \ref{fig:Cl_nu_feed}), while the intrinsic alignment parameter boosts it - if $A_\mathrm{IA}$ is negative - or damps it - if $A_\mathrm{IA}$ is positive - approximately in the same way (see for instance Figure 4 in Ref. \cite{Koh-KiDS+17}).
We would like to notice that this degeneracy appears only as long as we use a single resdhift bin, so performing tomography could help in alleviating or breaking it. In fact, while neutrinos affect all the redshift bins in  a similar way, intrinsic alignment depends much more on the source distribution and therefore has a different impact on different resdhift bins.
Moreover, although we do not show here the plots, another interesting point is the absence of degeneracy between the intrinsic alignment parameter and the other feedback parameters $\log M_\mathrm c$ and $\eta_\mathrm b$.
All combined, these results, limited to the analysis of matter 3D clustering, yield the conclusion that in the BCM model the measurement of neutrino mass will not be affected by baryon feedback nor by intrinsic alignment.

\section{Discussion and conclusions}
\label{sec:discussion}

In this paper we have shown that the effect of baryons on the matter and shear power spectra can be disentangled \textit{at the matter perturbation level} by using a tomographic analysis. To do so, we performed a likelihood analysis for the matter and shear power spectra using the Markov Chain Monte Carlo method.

In our analysis, we accounted for both statistical error, i.e. cosmic variance, as well as the systematic error affecting the theoretical model. In the perspective where all theoretical predictions are based on fitting functions or emulators based on numerical, N-body simulations, the systematic uncertainty reflects the limitations of this approach in providing an accurate description of the non-linear matter power spectrum. We adopted the formula by Ref. \cite{Bird-HALOFIT+12} to describe the scale-dependence of this systematic error, negligible at the largest, linear scales while growing monotonically until reaching a $\sim2\%$ level at scales of $0.5 \kMpc$  at $z=0$. In addition we follow Ref. \cite{Baldauf+16} to account for the correlation such error induces among different wavenumbers.

We investigated directly the effect of baryon feedback, another important source of systematic error, using the Baryon Correction Model by Ref. \cite{Schneider+15}.
Their fitting formula depends on three parameters ($M_c$, $\eta_\mathrm b$, $z_\mathrm c$) with well-established physical meanings. The main focus of our work has been the specific study of their degeneracies with neutrino masses.

As a first test, we explored the possibility that baryonic effects alone could mimic the characteristic suppression of power at small scales due to a non-vanishing neutrino mass. 
As expected, for both clustering and cosmic shear, we found that  growing values of the parameter $\logMc$, the mass below which halos are emptied from their gas, can reproduce, to some extent, the effect on an increasing neutrino mass.
However, the two probes, matter clustering and cosmic shear, interestingly prefer quite different values of this parameter, highlighting the relevance of their combination. In addition, for matter clustering in particular, the recovered value of the $\logMc$ parameter strongly depends on the maximum wavenumber included in the analysis. These differences can be up to $\sim 9 \sigma$ for $M_\nu = 0.45$ eV for the \textit{Euclid}-like observational set-up we considered, a clear hint that massive neutrinos and baryonic feedback effects will be indeed distinguishable in upcoming surveys.

As a second test, the central analysis carried out in this work, we studied directly the degeneracy between the two effects allowing the parameters of the baryonic feedback model and the neutrino mass to vary simultaneously.
We performed this Monte-Carlo analysis for 16 different sets of fiducial parameters in order to properly explore how such degeneracy depends on the assumed baryonic feedback model, a priori unknown.
In all cases, we were able to recover the input neutrino mass.
In particular, in the matter clustering case, the degeneracy between neutrino mass and feedback parameters is completely absent, while an interesting correlation exists in the plane $\log M_c - \eta_b$.
For the weak lensing shear power spectrum, the convolution of the matter power spectrum with the lensing kernel smooths-out the different scale-dependent features of the two effects leading to a noticeable degeneracy between the neutrino mass $M_\nu$ and the baryon feedback parameter $\logMc$. Despite this fact, the estimate of $M_\nu$ by cosmic shear  is still unbiased, with a maximum difference with respect to the true value of $\sim 0.25 \sigma$.

Finally we considered how these results are affected by the additional systematic represented by the intrinsic alignment effect in weak lensing survey.
We employed the linear alignment model by Ref. \cite{Hirata-IA+04} to introduce the intrinsic alignment effect on the shear spectra. For this part we fixed the $z_c$ parameter to a value of 2: this choice is motivated by the fact that in the previous part of the analysis we were never able to constrain it, as its effects are pretty small on the spectra and scales that we consider. The MCMC analysis was therefore performed with 4 free parameters: $M_\nu, \log M_c, \eta_b, A_\mathrm{IA}$.
Again, we were able to recover the right input values for what concerns neutrino mass and the $A_\mathrm{IA}$ parameter.
The posterior PDFs and contours for the cases with and without intrinsic alignment are almost identical (see Figure \ref{fig:triangle}), while we find a degeneracy pattern in the $M_\nu-A_\mathrm{IA}$ plane (Figure \ref{fig:Mnu_vs_AIA}) that is more pronounced when $A_\mathrm{IA}$ is negative.
Also, such parameter seems not to suffer from any degeneracy with the other feedback parameters.

In conclusion, if the BCM is used as a baryon feedback fiducial model, measurements on the neutrino mass from future surveys are likely not be affected by biases due to the degeneracy between neutrino masses and the feedback parameters.

Clearly, this is a first step in tackling a quite complex problem that requires much further work. Future developments concerning the clustering involve the introduction of a bias model and redshift-space distortions. Another interesting step would be to explore the degeneracies of the feedback parameters with the cosmological ones. Finally, more interesting insights (and hopefully some improvements in the constraints) could be obtained by studying the cross-correlation between the two observables, for which a detailed study of the cross-covariance matrix is needed.

\acknowledgments
The authors are supported by the INFN INDARK PD51 grant.


\bibliographystyle{unsrt}
\bibliography{bibtex}

\newcommand{\noopsort}[1]{}
\begin{thebibliography}{10}

\bibitem{DES_1yr+18}
T.~M.~C. {Abbott}, F.~B. {Abdalla}, et~al.
\newblock {The Dark Energy Survey Data Release 1}.
\newblock {\em ArXiv e-prints}, January 2018.

\bibitem{DES_lensing+17}
M.~A. {Troxel}, N.~{MacCrann}, J.~{Zuntz}, et~al.
\newblock {Dark Energy Survey Year 1 Results: Cosmological Constraints from
  Cosmic Shear}.
\newblock {\em ArXiv e-prints}, August 2017.

\bibitem{LP_massive_nu+06}
J.~{Lesgourgues} and S.~{Pastor}.
\newblock {Massive neutrinos and cosmology}.
\newblock {\em \physrep}, 429:307--379, July 2006.

\bibitem{Giusarma+16}
E.~{Giusarma}, M.~{Gerbino}, O.~{Mena}, S.~{Vagnozzi}, S.~{Ho}, and
  K.~{Freese}.
\newblock {Improvement of cosmological neutrino mass bounds}.
\newblock {\em \prd}, 94(8):083522, October 2016.

\bibitem{Beutler+14}
F.~{Beutler}, S.~{Saito}, J.~R. {Brownstein}, C.-H. {Chuang}, A.~J. {Cuesta},
  W.~J. {Percival}, A.~J. {Ross}, N.~P. {Ross}, D.~P. {Schneider},
  L.~{Samushia}, A.~G. {S{\'a}nchez}, H.-J. {Seo}, J.~L. {Tinker}, C.~{Wagner},
  and B.~A. {Weaver}.
\newblock {The clustering of galaxies in the SDSS-III Baryon Oscillation
  Spectroscopic Survey: signs of neutrino mass in current cosmological data
  sets}.
\newblock {\em \mnras}, 444:3501--3516, November 2014.

\bibitem{Battye+14}
R.~A. {Battye} and A.~{Moss}.
\newblock {Evidence for Massive Neutrinos from Cosmic Microwave Background and
  Lensing Observations}.
\newblock {\em Physical Review Letters}, 112(5):051303, February 2014.

\bibitem{DiValentino+17}
E.~{Di Valentino}, A.~{Melchiorri}, E.~V. {Linder}, and J.~{Silk}.
\newblock {Constraining dark energy dynamics in extended parameter space}.
\newblock {\em \prd}, 96(2):023523, July 2017.

\bibitem{Planck+15}
{Planck Collaboration}, P.~A.~R. {Ade}, N.~{Aghanim}, M.~{Arnaud},
  M.~{Ashdown}, J.~{Aumont}, C.~{Baccigalupi}, A.~J. {Banday}, R.~B.
  {Barreiro}, J.~G. {Bartlett}, et~al.
\newblock {Planck 2015 results. XIII. Cosmological parameters}.
\newblock {\em \aap}, 594:A13, September 2016.

\bibitem{PD-lya-nu+15}
N.~{Palanque-Delabrouille}, C.~{Y{\`e}che}, J.~{Baur}, C.~{Magneville},
  G.~{Rossi}, J.~{Lesgourgues}, A.~{Borde}, E.~{Burtin}, J.-M. {LeGoff},
  J.~{Rich}, M.~{Viel}, and D.~{Weinberg}.
\newblock {Neutrino masses and cosmology with Lyman-alpha forest power
  spectrum}.
\newblock {\em \jcap}, 11:011, November 2015.

\bibitem{OWLs+10}
J.~{Schaye}, C.~{Dalla Vecchia}, C.~M. {Booth}, R.~P.~C. {Wiersma},
  T.~{Theuns}, M.~R. {Haas}, S.~{Bertone}, A.~R. {Duffy}, I.~G. {McCarthy}, and
  F.~{van de Voort}.
\newblock {The physics driving the cosmic star formation history}.
\newblock {\em \mnras}, 402:1536--1560, March 2010.

\bibitem{Van_Daalen-feedback+11}
M.~P. {van Daalen}, J.~{Schaye}, C.~M. {Booth}, and C.~{Dalla Vecchia}.
\newblock {The effects of galaxy formation on the matter power spectrum: a
  challenge for precision cosmology}.
\newblock {\em \mnras}, 415:3649--3665, August 2011.

\bibitem{Semboloni+11}
E.~{Semboloni}, H.~{Hoekstra}, J.~{Schaye}, M.~P. {van Daalen}, and I.~G.
  {McCarthy}.
\newblock {Quantifying the effect of baryon physics on weak lensing
  tomography}.
\newblock {\em \mnras}, 417:2020--2035, November 2011.

\bibitem{Mummery-fb_vs_nu+17}
B.~O. {Mummery}, I.~G. {McCarthy}, S.~{Bird}, and J.~{Schaye}.
\newblock {The separate and combined effects of baryon physics and neutrino
  free streaming on large-scale structure}.
\newblock {\em \mnras}, 471:227--242, October 2017.

\bibitem{Sun+09}
M.~{Sun}, G.~M. {Voit}, M.~{Donahue}, C.~{Jones}, W.~{Forman}, and
  A.~{Vikhlinin}.
\newblock {Chandra Studies of the X-Ray Gas Properties of Galaxy Groups}.
\newblock {\em \apj}, 693:1142--1172, March 2009.

\bibitem{Lovisari+15}
L.~{Lovisari}, T.~H. {Reiprich}, and G.~{Schellenberger}.
\newblock {Scaling properties of a complete X-ray selected galaxy group
  sample}.
\newblock {\em \aap}, 573:A118, January 2015.

\bibitem{Eckert+16}
D.~{Eckert}, S.~{Ettori}, J.~{Coupon}, F.~{Gastaldello}, M.~{Pierre}, J.-B.
  {Melin}, A.~M.~C. {Le Brun}, I.~G. {McCarthy}, C.~{Adami}, L.~{Chiappetti},
  L.~{Faccioli}, P.~{Giles}, S.~{Lavoie}, J.~P. {Lef{\`e}vre}, M.~{Lieu},
  A.~{Mantz}, B.~{Maughan}, S.~{McGee}, F.~{Pacaud}, S.~{Paltani},
  T.~{Sadibekova}, G.~P. {Smith}, and F.~{Ziparo}.
\newblock {The XXL Survey. XIII. Baryon content of the bright cluster sample}.
\newblock {\em \aap}, 592:A12, June 2016.

\bibitem{BAHAMAS+17}
I.~G. {McCarthy}, J.~{Schaye}, S.~{Bird}, and A.~M.~C. {Le Brun}.
\newblock {The BAHAMAS project: calibrated hydrodynamical simulations for
  large-scale structure cosmology}.
\newblock {\em \mnras}, 465:2936--2965, March 2017.

\bibitem{FABLE+18}
N.~A. {Henden}, E.~{Puchwein}, S.~{Shen}, and D.~{Sijacki}.
\newblock {The FABLE simulations: a feedback model for galaxies, groups, and
  clusters}.
\newblock {\em \mnras}, 479:5385--5412, October 2018.

\bibitem{Schneider+15}
A.~{Schneider} and R.~{Teyssier}.
\newblock {A new method to quantify the effects of baryons on the matter power
  spectrum}.
\newblock {\em \jcap}, 12:049, December 2015.

\bibitem{HD_feedback+15}
J.~{Harnois-D{\'e}raps}, L.~{van Waerbeke}, M.~{Viola}, and C.~{Heymans}.
\newblock {Baryons, neutrinos, feedback and weak gravitational lensing}.
\newblock {\em \mnras}, 450:1212--1223, June 2015.

\bibitem{Chisari-feedback+18}
N.~E. {Chisari}, M.~L.~A. {Richardson}, J.~{Devriendt}, Y.~{Dubois},
  A.~{Schneider}, A.~M.~C. {Le Brun}, R.~S. {Beckmann}, S.~{Peirani},
  A.~{Slyz}, and C.~{Pichon}.
\newblock {The impact of baryons on the matter power spectrum from the
  Horizon-AGN cosmological hydrodynamical simulation}.
\newblock {\em ArXiv e-prints}, January 2018.

\bibitem{Mead-b+15}
A.~J. {Mead}, J.~A. {Peacock}, C.~{Heymans}, S.~{Joudaki}, and A.~F. {Heavens}.
\newblock {An accurate halo model for fitting non-linear cosmological power
  spectra and baryonic feedback models}.
\newblock {\em \mnras}, 454:1958--1975, December 2015.

\bibitem{Koh-KiDS+17}
F.~{K{\"o}hlinger}, M.~{Viola}, B.~{Joachimi}, H.~{Hoekstra}, et~al.
\newblock {KiDS-450: the tomographic weak lensing power spectrum and
  constraints on cosmological parameters}.
\newblock {\em \mnras}, 471:4412--4435, November 2017.

\bibitem{Audren+12}
B.~{Audren}, J.~{Lesgourgues}, S.~{Bird}, M.~G. {Haehnelt}, and M.~{Viel}.
\newblock {Neutrino masses and cosmological parameters from a Euclid-like
  survey: Markov Chain Monte Carlo forecasts including theoretical errors}.
\newblock {\em \jcap}, page 026, January 2013.

\bibitem{Sprenger+18}
T.~{Sprenger}, M.~{Archidiacono}, T.~{Brinckmann}, S.~{Clesse}, and
  J.~{Lesgourgues}.
\newblock {Cosmology in the era of Euclid and the Square Kilometre Array}.
\newblock {\em ArXiv e-prints}, January 2018.

\bibitem{Ichiki+11}
K.~{Ichiki} and M.~{Takada}.
\newblock {Impact of massive neutrinos on the abundance of massive clusters}.
\newblock {\em \prd}, 85(6):063521, March 2012.

\bibitem{nuLCDM1+13}
F.~{Villaescusa-Navarro}, F.~{Marulli}, M.~{Viel}, E.~{Branchini},
  E.~{Castorina}, E.~{Sefusatti}, and S.~{Saito}.
\newblock {Cosmology with massive neutrinos I: towards a realistic modeling of
  the relation between matter, haloes and galaxies}.
\newblock {\em \jcap}, 3:011, March 2014.

\bibitem{nuLCDM2+13}
E.~{Castorina}, E.~{Sefusatti}, R.~K. {Sheth}, F.~{Villaescusa-Navarro}, and
  M.~{Viel}.
\newblock {Cosmology with massive neutrinos II: on the universality of the halo
  mass function and bias}.
\newblock {\em \jcap}, 2:049, February 2014.

\bibitem{nuLCDM3+13}
M.~{Costanzi}, F.~{Villaescusa-Navarro}, M.~{Viel}, J.-Q. {Xia}, S.~{Borgani},
  E.~{Castorina}, and E.~{Sefusatti}.
\newblock {Cosmology with massive neutrinos III: the halo mass function and an
  application to galaxy clusters}.
\newblock {\em \jcap}, 12:012, December 2013.

\bibitem{Takada_WL+04}
M.~{Takada} and B.~{Jain}.
\newblock {Cosmological parameters from lensing power spectrum and bispectrum
  tomography}.
\newblock {\em \mnras}, 348:897--915, March 2004.

\bibitem{Lensing_is_low+16}
A.~{Leauthaud}, S.~{Saito}, S.~{Hilbert}, et~al.
\newblock {Lensing is low: cosmology, galaxy formation or new physics?}
\newblock {\em \mnras}, 467:3024--3047, May 2017.

\bibitem{Problems_KiDS+17}
G.~{Efstathiou} and P.~{Lemos}.
\newblock {Statistical inconsistencies in the KiDS-450 data set}.
\newblock {\em \mnras}, 476:151--157, May 2018.

\bibitem{McCarthy+17}
I.~G. {McCarthy}, S.~{Bird}, J.~{Schaye}, J.~{Harnois-Deraps}, A.~S. {Font},
  and L.~{van Waerbeke}.
\newblock {The BAHAMAS project: the CMB-large-scale structure tension and the
  roles of massive neutrinos and galaxy formation}.
\newblock {\em \mnras}, 476:2999--3030, May 2018.

\bibitem{Mead-nu+16}
A.~J. {Mead}, C.~{Heymans}, L.~{Lombriser}, J.~A. {Peacock}, O.~I. {Steele},
  and H.~A. {Winther}.
\newblock {Accurate halo-model matter power spectra with dark energy, massive
  neutrinos and modified gravitational forces}.
\newblock {\em \mnras}, 459:1468--1488, June 2016.

\bibitem{DEMNUni+16}
E.~{Castorina}, C.~{Carbone}, J.~{Bel}, E.~{Sefusatti}, and K.~{Dolag}.
\newblock {DEMNUni: the clustering of large-scale structures in the presence of
  massive neutrinos}.
\newblock {\em \jcap}, 7:043, July 2015.

\bibitem{Paco-Neutrinos+17}
F.~{Villaescusa-Navarro}, A.~{Banerjee}, N.~{Dalal}, E.~{Castorina},
  R.~{Scoccimarro}, R.~{Angulo}, and D.~N. {Spergel}.
\newblock {The imprint of neutrinos on clustering in redshift-space}.
\newblock {\em ArXiv e-prints}, August 2017.

\bibitem{Ruggeri+18}
R.~{Ruggeri}, E.~{Castorina}, C.~{Carbone}, and E.~{Sefusatti}.
\newblock {DEMNUni: massive neutrinos and the bispectrum of large scale
  structures}.
\newblock {\em \jcap}, 3:003, March 2018.

\bibitem{Bartelmann_WL+01}
M.~{Bartelmann} and P.~{Schneider}.
\newblock {Weak gravitational lensing}.
\newblock {\em \physrep}, 340:291--472, January 2001.

\bibitem{Hoek_WL+08}
H.~{Hoekstra} and B.~{Jain}.
\newblock {Weak Gravitational Lensing and Its Cosmological Applications}.
\newblock {\em Annual Review of Nuclear and Particle Science}, 58:99--123,
  November 2008.

\bibitem{Kilbinger-full-sky+17}
M.~{Kilbinger}, C.~{Heymans}, M.~{Asgari}, S.~{Joudaki}, P.~{Schneider},
  P.~{Simon}, L.~{Van Waerbeke}, J.~{Harnois-D{\'e}raps}, H.~{Hildebrandt},
  F.~{K{\"o}hlinger}, K.~{Kuijken}, and M.~{Viola}.
\newblock {Precision calculations of the cosmic shear power spectrum
  projection}.
\newblock {\em \mnras}, 472:2126--2141, December 2017.

\bibitem{Bird-HALOFIT+12}
S.~{Bird}, M.~{Viel}, and M.~G. {Haehnelt}.
\newblock {Massive neutrinos and the non-linear matter power spectrum}.
\newblock {\em \mnras}, 420:2551--2561, March 2012.

\bibitem{brandbyge08}
J.~{Brandbyge} and S.~{Hannestad}.
\newblock {Grid based linear neutrino perturbations in cosmological N-body
  simulations}.
\newblock {\em \jcap}, 5:002, May 2009.

\bibitem{vielhaehneltspringel2010}
M.~{Viel}, M.~G. {Haehnelt}, and V.~{Springel}.
\newblock {The effect of neutrinos on the matter distribution as probed by the
  intergalactic medium}.
\newblock {\em \jcap}, 6:015, June 2010.

\bibitem{Jing+05}
Y.~P. {Jing}, P.~{Zhang}, W.~P. {Lin}, L.~{Gao}, and V.~{Springel}.
\newblock {The Influence of Baryons on the Clustering of Matter and
  Weak-Lensing Surveys}.
\newblock {\em \apjl}, 640:L119--L122, April 2006.

\bibitem{Perotto+06}
L.~{Perotto}, J.~{Lesgourgues}, S.~{Hannestad}, H.~{Tu}, and Y.~{Y Y Wong}.
\newblock {Probing cosmological parameters with the CMB: forecasts from Monte
  Carlo simulations}.
\newblock {\em \jcap}, 10:013, October 2006.

\bibitem{Camb}
A.~{Lewis} and A.~{Challinor}.
\newblock {CAMB: Code for Anisotropies in the Microwave Background}.
\newblock Astrophysics Source Code Library, February 2011.

\bibitem{Bispectrum-Sefusatti+06}
E.~{Sefusatti}, M.~{Crocce}, S.~{Pueblas}, and R.~{Scoccimarro}.
\newblock {Cosmology and the bispectrum}.
\newblock {\em \prd}, 74(2):023522, July 2006.

\bibitem{Covariance-Hamilton+05}
A.~J.~S. {Hamilton}, C.~D. {Rimes}, and R.~{Scoccimarro}.
\newblock {On measuring the covariance matrix of the non-linear power spectrum
  from simulations}.
\newblock {\em \mnras}, 371:1188--1204, September 2006.

\bibitem{Super_sample_covariance-Takada+13}
M.~{Takada} and W.~{Hu}.
\newblock {Power spectrum super-sample covariance}.
\newblock {\em \prd}, 87(12):123504, June 2013.

\bibitem{Precision_Pk-Schneider+16}
A.~{Schneider}, R.~{Teyssier}, D.~{Potter}, J.~{Stadel}, J.~{Onions}, D.~S.
  {Reed}, R.~E. {Smith}, V.~{Springel}, F.~R. {Pearce}, and R.~{Scoccimarro}.
\newblock {Matter power spectrum and the challenge of percent accuracy}.
\newblock {\em \jcap}, 4:047, April 2016.

\bibitem{Zennaro+17}
M.~{Zennaro}, J.~{Bel}, F.~{Villaescusa-Navarro}, C.~{Carbone}, E.~{Sefusatti},
  and L.~{Guzzo}.
\newblock {Initial conditions for accurate N-body simulations of massive
  neutrino cosmologies}.
\newblock {\em \mnras}, 466:3244--3258, April 2017.

\bibitem{Halofit-Smith+03}
R.~E. {Smith}, J.~A. {Peacock}, A.~{Jenkins}, S.~D.~M. {White}, C.~S. {Frenk},
  F.~R. {Pearce}, P.~A. {Thomas}, G.~{Efstathiou}, and H.~M.~P. {Couchman}.
\newblock {Stable clustering, the halo model and non-linear cosmological power
  spectra}.
\newblock {\em \mnras}, 341:1311--1332, June 2003.

\bibitem{Takahashi+12}
R.~{Takahashi}, M.~{Sato}, T.~{Nishimichi}, A.~{Taruya}, and M.~{Oguri}.
\newblock {Revising the Halofit Model for the Nonlinear Matter Power Spectrum}.
\newblock {\em \apj}, 761:152, December 2012.

\bibitem{CoyoteI}
K.~{Heitmann}, M.~{White}, C.~{Wagner}, S.~{Habib}, and D.~{Higdon}.
\newblock {The Coyote Universe. I. Precision Determination of the Nonlinear
  Matter Power Spectrum}.
\newblock {\em \apj}, 715:104--121, May 2010.

\bibitem{CoyoteII}
K.~{Heitmann}, D.~{Higdon}, M.~{White}, S.~{Habib}, B.~J. {Williams},
  E.~{Lawrence}, and C.~{Wagner}.
\newblock {The Coyote Universe. II. Cosmological Models and Precision Emulation
  of the Nonlinear Matter Power Spectrum}.
\newblock {\em \apj}, 705:156--174, November 2009.

\bibitem{CoyoteIII}
E.~{Lawrence}, K.~{Heitmann}, M.~{White}, D.~{Higdon}, C.~{Wagner}, S.~{Habib},
  and B.~{Williams}.
\newblock {The Coyote Universe. III. Simulation Suite and Precision Emulator
  for the Nonlinear Matter Power Spectrum}.
\newblock {\em \apj}, 713:1322--1331, April 2010.

\bibitem{Cosmic_emu+10}
E.~{Lawrence}, K.~{Heitmann}, M.~{White}, D.~{Higdon}, C.~{Wagner}, S.~{Habib},
  and B.~{Williams}.
\newblock {CosmicEmu: Cosmic Emulator for the Dark Matter Power Spectrum}.
\newblock Astrophysics Source Code Library, October 2010.

\bibitem{EuclidEmulator}
{Euclid Collaboration}, M.~{Knabenhans}, J.~{Stadel}, S.~{Marelli},
  D.~{Potter}, R.~{Teyssier}, L.~{Legrand}, A.~{Schneider}, B.~{Sudret},
  L.~{Blot}, S.~{Awan}, C.~{Burigana}, C.~S. {Carvalho}, H.~{Kurki-Suonio}, and
  G.~{Sirri}.
\newblock {Euclid preparation: II. The EuclidEmulator -- A tool to compute the
  cosmology dependence of the nonlinear matter power spectrum}.
\newblock {\em ArXiv e-prints}, September 2018.

\bibitem{Smith-Angulo+18}
R.~E. {Smith} and R.~E. {Angulo}.
\newblock {Precision modelling of the matter power spectrum in a Planck-like
  Universe}.
\newblock {\em ArXiv e-prints}, June 2018.

\bibitem{Baldauf+16}
T.~{Baldauf}, M.~{Mirbabayi}, M.~{Simonovi{\'c}}, and M.~{Zaldarriaga}.
\newblock {LSS constraints with controlled theoretical uncertainties}.
\newblock {\em ArXiv e-prints}, February 2016.

\bibitem{Troxel-IA+15}
M.~A. {Troxel} and M.~{Ishak}.
\newblock {The intrinsic alignment of galaxies and its impact on weak
  gravitational lensing in an era of precision cosmology}.
\newblock {\em \physrep}, 558:1--59, February 2015.

\bibitem{Heavens-IA+00}
A.~{Heavens}, A.~{Refregier}, and C.~{Heymans}.
\newblock {Intrinsic correlation of galaxy shapes: implications for weak
  lensing measurements}.
\newblock {\em \mnras}, 319:649--656, December 2000.

\bibitem{Croft-IA+00}
R.~A.~C. {Croft} and C.~A. {Metzler}.
\newblock {Weak-Lensing Surveys and the Intrinsic Correlation of Galaxy
  Ellipticities}.
\newblock {\em \apj}, 545:561--571, December 2000.

\bibitem{Hirata-IA+04}
C.~M. {Hirata} and U.~{Seljak}.
\newblock {Intrinsic alignment-lensing interference as a contaminant of cosmic
  shear}.
\newblock {\em \prd}, 70(6):063526, September 2004.

\end{thebibliography}

\end{document}